\RequirePackage{lineno} 

\documentclass[twocolumn,showpacs,preprintnumbers,amsmath,amssymb,nofootinbib]{revtex4}

\usepackage{graphicx}
\usepackage{dcolumn}
\usepackage{bm}

\usepackage{epstopdf}
\def\bea{\begin{eqnarray}}
\def\eea{\end{eqnarray}}

\def\pp{\mbox{$p$-$p$}}
\def\pa{\mbox{$p$-$A$}}

\def\auau{\mbox{Au-Au}}

\def\pbpb{\mbox{Pb-Pb}}
\def\aa{\mbox{$A$-$A$}}
\def\nn{\mbox{$N$-$N$}}

\def\dau{\mbox{$d$-Au}}

\def\pt{$p_t$}
\def\titlept{$\bf p_t$}
\def\titlepp{$\bf p$-$\bf p$}
\def\mt{$m_t$}
\def\yt{$y_t$}

\def\nch{$n_{ch}$}

\def\ppb{\mbox{$p$-Pb}}
\def\pn{\mbox{$p$-N}}

\begin{document} 

\setlength{\pdfpagewidth}{8.5in}
\setlength{\pdfpageheight}{11in}

\setpagewiselinenumbers
\modulolinenumbers[5]

\preprint{Version 1.4}

\title{
Statistical evaluation of fitted models applied to $\bf p_t$ spectrum data\\ from 5 TeV and 13 TeV $\bf p$-$\bf p$ collisions at the large hadron collider
}

\author{Thomas A.\ Trainor}\affiliation{CENPA 354290, University of Washington, Seattle, Washington 98195}


\date{\today}

\begin{abstract}

In a recent  analysis of high-statistics \pt\ spectra from 5 and 13 TeV $p$-$p$ collisions a two-component (soft+hard) model (TCM) of hadron production near midrapidity, based on fixed model functions independent of collision multiplicity $n_{ch}$ or event selection criteria, was employed as a reference to determine data biases induced by the selection criteria. As in previous studies the fixed TCM accurately isolated jet-related and nonjet contributions to spectra. It was observed that two selection criteria (V0M and SPD), based on different pseudorapidity $\eta$ acceptances, bias the spectrum hard component (jet fragment distribution) in different but complementary ways whereas the soft component is not significantly biased. In the present study TCM model functions are adjusted (i.e.\ fitted) to accommodate data. The effect of selection bias is then represented by smooth evolution of model parameters with $n_{ch}$. To evaluate the quality of model fits, Z-scores (data-model {\em differences} divided by statistical uncertainties) are evaluated and compared with data/model {\em ratios} that are often used for such comparisons. Based on Z-scores the variable TCM is compared with two other frequently-invoked models: the Tsallis model and the blast-wave model as they have been recently applied to the same spectrum data. The results are relevant to recent claims that collectivity (various flow manifestations) as a possible manifestation of QGP formation is observed in small systems. Data systematics suggest that these $p$-$p$ $p_t$ spectra are consistent with conventional QCD.

\end{abstract}


\maketitle

\section{Introduction} \label{intro}

The ALICE collaboration recently published a comprehensive high-statistics study of \pt\ spectra from 5 TeV and 13 TeV \pp\ collisions~\cite{alicenewspec}. The study employed two methods (V0M and SPD) to sort collision events into ten multiplicity classes each and application of {\em spherocity} $S_0$, a measure of the azimuthal asymmetry of distributed $\vec p_t$, to estimate the ``jettiness'' of events. Several methods were applied to determine variation of spectrum shape with charge multiplicity, event selection method and spherocity. A stated goal was ``...to investigate the importance of jets in high-multiplicity pp collisions and their contribution to charged-particle production at low $p_T$.'' While Ref.~\cite{alicenewspec} described the relation of three Monte Carlo versions to various data features there was no conclusion about jet contributions to spectra, whereas  the contribution of minimum-bias (MB) dijets to \pt\ spectra over the entire detector \pt\ acceptance has been established over more than a decade based on the two-component model of hadron production in A-B collisions~\cite{ppprd,hardspec,ppquad,alicetomspec,ppbpid,newpptcm}.

The two-component (soft + hard) model (TCM) of hadron production near mid-rapidity was initially derived from the charge-multiplicity \nch\ dependence of \pt\ spectra from 200 GeV \pp\ collisions~\cite{ppprd}. The \nch\ dependence of yields, spectra and two-particle correlations has played a key role in establishing (a) the nature of hadron production mechanisms in \pp\ collisions and (b) that the TCM hard component of \pt\ spectra is {\em quantitatively} consistent with predictions based on event-wise reconstructed jets~\cite{fragevo,jetspec,jetspec2}. The question of recently claimed collectivity or flows in small (\pp\ and \pa) systems has been addressed on the basis of the resolved TCM soft and hard components and evidence (or not) for radial flow in differential studies of \pt\ spectra~\cite{hardspec,ppquad}.

Reference~\cite{newpptcm} responded to spectrum data reported in Ref.~\cite{alicenewspec} by applying a {\em fixed} TCM to investigate spectrum structure. The term ``fixed'' is meant to indicate that individual TCM model functions are independent of event charge multiplicity \nch\ or event selection method. Given that approach the TCM serves as a {\em predictive reference} that enables {\em quantitative} assessment of the contribution of MB dijets to spectra over the complete \pt\ acceptance. Various selection biases arising from V0M, SPD and $S_0$ event selection are then precisely analyzed and their relation to jet production mechanisms are evaluated.

Reference~\cite{alicenewspec} includes multiple references to a currently-popular theme -- that certain data features observed in high-multiplicity \pp\ collisions appear similar to features observed in \aa\ collisions that are conventionally invoked to demonstrate achievement of QGP formation. Referring to similarities between data features in \aa\ and higher-\nch\ \pp\ collisions Ref.~\cite{alicenewspec} states that ``...radial and anisotropic [e.g.\ elliptic] flow, as well as strangeness enhancement, {\em are also observed} in pp and p-A collisions...[emphasis added].'' Such statements are based on certain preferred statistical analysis methods and   questionable fitting exercises with model functions. 

For example, a recent analysis of Ref.~\cite{alicenewspec} data based on model fits to spectra motivated by claims of collectivity in small systems is reported in Ref.~\cite{cleymans}. The two invoked models assume the context of a flowing dense medium created in \pp\ collisions. The Tsallis model is associated with a partially-equilibrated thermodynamic system and is used to infer ``kinetic freezeout'' parameters from \pt\ spectra. A blast-wave model is used to infer the mean transverse speed of a radially-expanding particle source presumably responding to (large?) density gradients.

In the present study  emphasis is placed on comparison of several models fitted to spectrum data and interpretation of model-fitting results. The quality of model descriptions is established by the {\em Z-score}, a standard statistical measure of data-model agreement. Within that context a variable TCM is introduced as a ``fitted'' spectrum model: $\hat S_0(y_t)$ and $\hat H_0(y_t)$ model parameters are adjusted within some limits to accommodate spectrum data. The quality of Tsallis model fits from Ref.~\cite{cleymans} is evaluated via similar methods pertaining to statistical significance. The blast-wave model is compared to identified-hadron spectra from 5 TeV \ppb\ collisions and corresponding TCM~\cite{ppbpid} as a precise test of radial-flow relevance. Two issues emerge: how do nonjet models (e.g.\ Tsallis and blast-wave) relate to jet-related spectrum structure, and what is the quality (via Z-scores) of the overall data description for any spectrum model?

This article is arranged as follows:
Section~\ref{pptcm} reviews the usual fixed TCM (soft and hard model functions unchanging) previously applied to spectrum data.
Section~\ref{tcmfit} describes an updated version of the TCM in which the two model functions are varied to accommodate spectrum data as a fitting exercise.
Section~\ref{tsallismodel} evaluates the quality of Tsallis model fits to V0M data.
Section~\ref{bwmodel} considers blast-wave (hydrodynamic)  model descriptions of hadron spectra in comparison with a fixed TCM applied to identified-hadron spectra from \ppb\ collisions, especially regarding evidence or not for radial flow.
Section~\ref{syserr} discusses systematic uncertainties.
Sections~\ref{disc} and~\ref{summ}  present discussion and summary.

\section{\titlepp\ \titlept\ spectrum fixed TCM} \label{pptcm}

The \pt\ spectrum TCM, first reported for 200 GeV \pp\ collisions in Ref.~\cite{ppprd}, is an accurate description of yields, spectra and two-particle correlations for A-B collision systems based on linear superposition of \nn\ or parton-parton collisions. In general, the fixed TCM serves as a {\em predictive reference} for any collision system. Deviations from the TCM then provide systematic and quantitative information on details of collision mechanisms. In this section the basic spectrum TCM is reviewed and then applied to 5 TeV and 13 TeV \pt\ spectra for two event selection methods from Ref.~\cite{alicenewspec}.  This introductory material was presented previously in Ref.~\cite{newpptcm}.  In the next section the basic TCM is modified to accommodate different multiplicity \nch\ classes and event selection methods.

\subsection{\titlept\ spectrum TCM for unidentified hadrons} \label{unidspec}

The \pt\ or \yt\ spectrum TCM is by definition the sum of soft and hard components with details inferred from data (e.g.\ Ref.~\cite{ppprd}). For \pp\ collisions the model is defined by
\bea  \label{rhotcm}
\bar \rho_{0}(y_t;n_{ch}) &\approx& \frac{d^2n_{ch}}{y_t dy_t d\eta}
\\ \nonumber
&=& \bar \rho_{s}(n_{ch}) \hat S_{0}(y_t) + \bar \rho_{h}(n_{ch}) \hat H_{0}(y_t),
\eea
where \nch\ is an event-class index and factorization of the dependences on \yt\ and \nch\ is a central feature of the spectrum TCM inferred from 200 GeV \pp\ spectrum data in Ref.~\cite{ppprd}. The motivation for transverse rapidity $y_{ti} \equiv \ln[(p_t + m_{ti})/m_i]$ (applied to hadron species $i$) is explained in Sec.~\ref{tcmmodel}. The \yt\ integral of Eq.~(\ref{rhotcm}) is $\bar \rho_0 \equiv n_{ch} / \Delta \eta = \bar \rho_s + \bar \rho_h$, a sum of soft and hard charge densities. $\hat S_{0}(y_t)$ and $\hat H_{0}(y_t)$ are unit-normal model functions independent of \nch. The centrally-important relation $\bar \rho_{h} \approx \alpha \bar \rho_{s}^2$ with $\alpha = O(0.01)$ is inferred from \pp\ spectrum data~\cite{ppprd,ppquad,alicetomspec}. $\bar \rho_s$ is then obtained from measured $\bar \rho_0$ as the root of the quadratic equation $\bar \rho_0  = \bar \rho_s + \alpha \bar \rho_s^2$ with  $\alpha$ determined by a systematic energy dependence derived from data trends covering a large energy interval~\cite{alicetomspec}. It is important to distinguish TCM model elements from spectrum data soft and hard components. It is useful to recall that \yt\ values 1, 2, 3, 4 and 5 are approximately equivalent to \pt\ values 0.15, 0.5, 1.4, 3.8 and 10 GeV/c.

\subsection{\titlept\ spectrum TCM model functions} \label{tcmmodel}

The \pp\ \pt\ spectrum soft component is most efficiently described on transverse mass \mt\ whereas the spectrum hard component is most efficiently described on transverse rapidity \yt. The spectrum TCM thus requires a heterogeneous set of variables for its simplest definition. The components can be easily transformed from one variable to the other by Jacobian factors defined below.

Given spectrum data in the form of Eq.~(\ref{rhotcm}) the unit-normal spectrum soft-component model $\hat S_0(y_t)$ is defined as the asymptotic limit of data spectra normalized in the form $X(y_t) \equiv \bar \rho_{0}(y_t;n_{ch}) / \bar \rho_s$ as \nch\ goes to zero. Hard components of data spectra are then defined as complementary to soft components, with the explicit form
\bea \label{yyt}
Y(y_t) \equiv \frac{1}{\alpha \bar \rho_s} \left[ X(y_t) - \hat S_0(y_t) \right],
\eea
directly comparable with TCM model function $\hat H_0(y_t)$.

The data soft component for a specific hadron species $i$ is typically well described by a L\'evy distribution on $m_{ti}  = \sqrt{p_t^2 + m_i^2}$. The unit-integral soft-component model is 
\bea \label{s00}
\hat S_{0i}(m_{ti}) &=& \frac{A_i}{[1 + (m_{ti} - m_i) / n_i T_i]^{n_i}},
\eea
where $m_{ti}$ is the transverse mass-energy for hadron species $i$ with mass $m_i$, $n_i$ is the L\'evy exponent, $T_i$ is the slope parameter and coefficient $A_i$ is determined by the unit-normal condition. Model parameters $(T_i,n_i)$ for several species of identified hadrons have been inferred from 5 TeV \ppb\ spectrum data as described in Ref.~\cite{ppbpid}. A soft-component model function for unidentified hadrons is defined as the weighted sum
\bea \label{s0m}
\hat S_{0}(m_{t}) &=& \sum_i z_{0i} \hat S_{0i}(m_{ti}),
\eea
where the weights for charged hadrons follow $\sum_i z_{0i} = 1$. 

The unit-normal hard-component model  is a Gaussian on $y_{t\pi} \equiv \ln[(p_t + m_{t\pi})/m_\pi]$ (as explained below) with exponential (on $y_t$) or power-law (on $p_t$) tail for larger \yt\
\bea \label{h00}
\hat H_{0}(y_t) &\approx & A \exp\left\{ - \frac{(y_t - \bar y_t)^2}{2 \sigma^2_{y_t}}\right\}~~~\text{near mode $\bar y_t$}
\\ \nonumber
&\propto &  \exp(- q y_t)~~~\text{for larger $y_t$ -- the tail},
\eea
where the transition from Gaussian to exponential on \yt\ is determined by slope matching~\cite{fragevo}. The $\hat H_0$ tail density on \pt\ varies approximately as power law $1/p_t^{q + 1.8} \approx  1/p_t^n$. Coefficient $A$ is determined by the unit-normal condition. 

All spectra are plotted vs pion rapidity $y_{t\pi}$ with pion mass assumed. The motivation is comparison of spectrum hard components demonstrated to arise from a common underlying jet spectrum on \pt~\cite{fragevo}, in which case $y_{t\pi}$ serves simply as a logarithmic measure of hadron \pt\ with well-defined zero. $\hat S_0(m_{t})$ in Eq.~(\ref{s0m}) is transformed to $y_{t\pi}$ via the Jacobian factor $m_{t\pi} p_t / y_{t\pi}$ to form  $\hat S_0(y_{t\pi})$ for unidentified hadrons. $\hat H_0(y_{t})$ in Eq.~(\ref{h00}) is always defined on $y_{t\pi}$ as noted. In general, plotting spectra on a logarithmic rapidity variable provides much-improved access to important spectrum structure in the low-\pt\ interval {\em where the majority of jet fragments appear}.

\subsection{\titlepp\ \titlept\ spectrum data} \label{tcmspecdat}

\begin{figure}[h]
	\includegraphics[width=3.3in,height=1.6in]{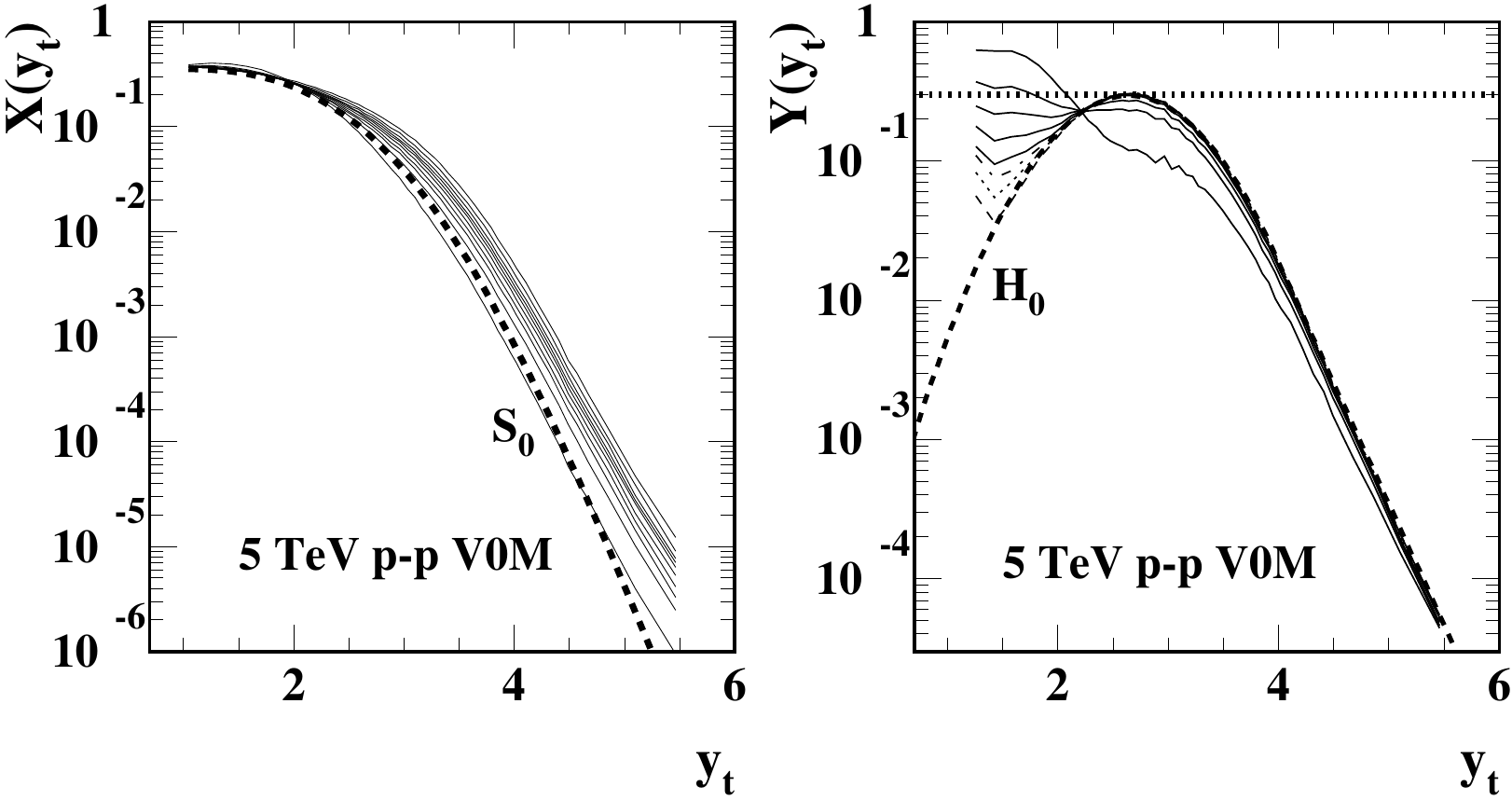}
	\put(-140,89) {\bf (a)}
	\put(-24,89) {\bf (b)}\\
	\includegraphics[width=3.3in,height=1.6in]{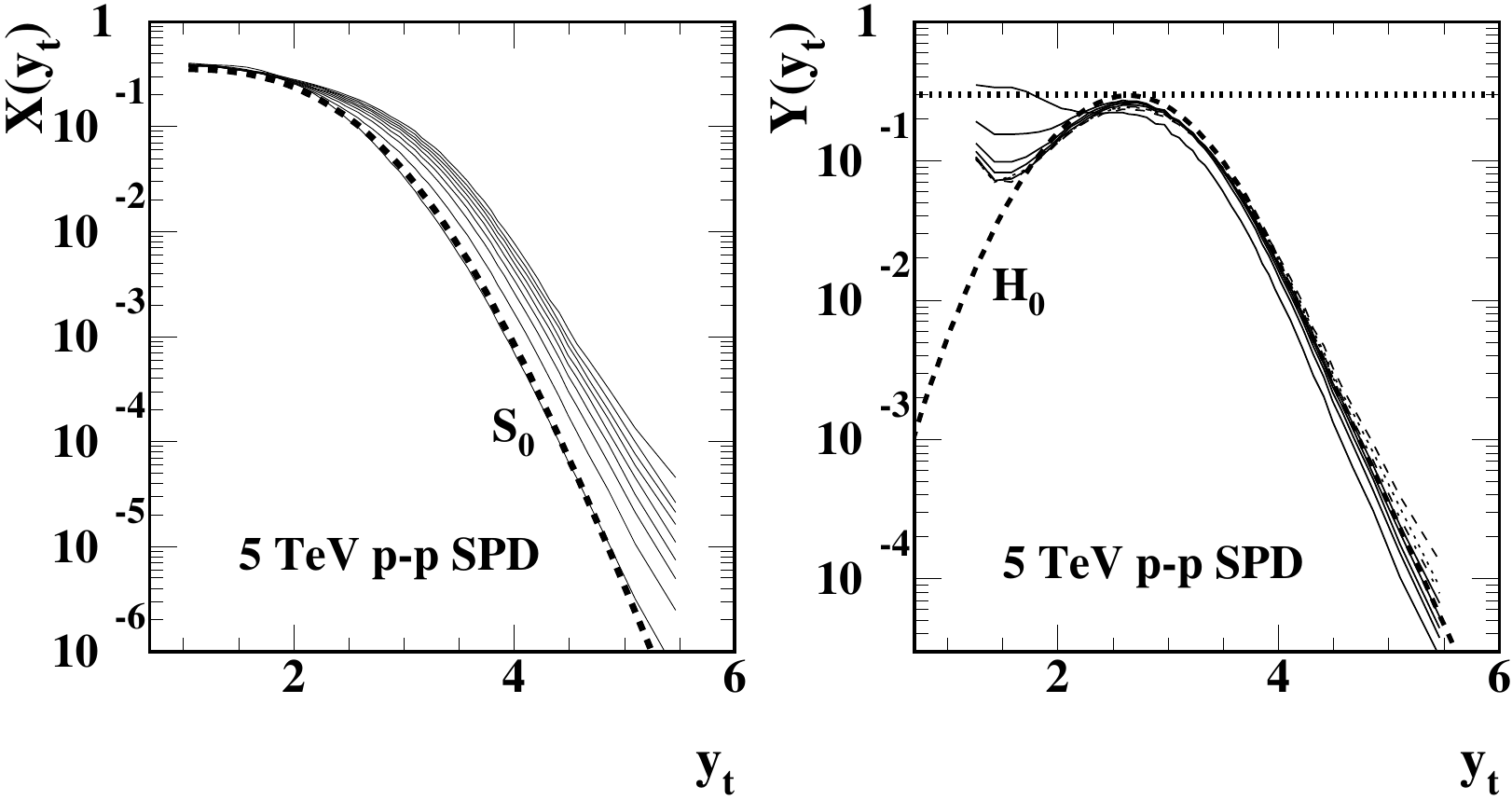}
	\put(-140,89) {\bf (c)}
	\put(-24,89) {\bf (d)}
	\caption{\label{tcm5}
		Left: Normalized \yt\ spectra in the form $X(y_t)$ from ten (V0M) or nine (SPD) multiplicity classes of 5 TeV \pp\ collisions for V0M (a) and SPD (c) event selection.
		Right: Normalized spectrum hard components in the form $Y(y_t)$ for data in the left panels for  V0M (b) and SPD (d) event selection. The bold dashed curves are TCM model functions.
	} 
\end{figure}

\begin{figure}[h]
	\includegraphics[width=3.3in,height=1.6in]{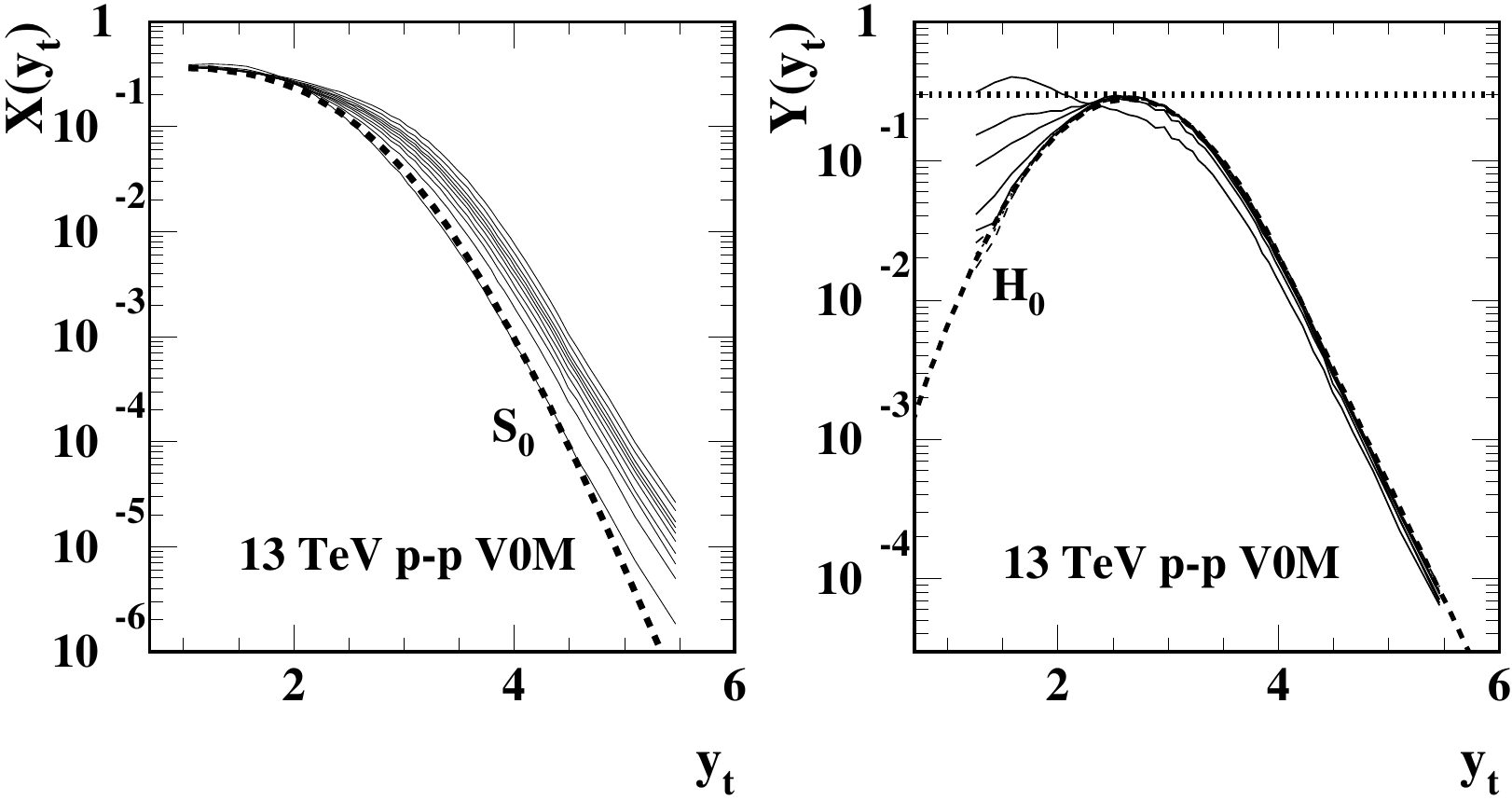}
	\put(-140,89) {\bf (a)}
	\put(-24,89) {\bf (b)}\\
	\includegraphics[width=3.3in,height=1.6in]{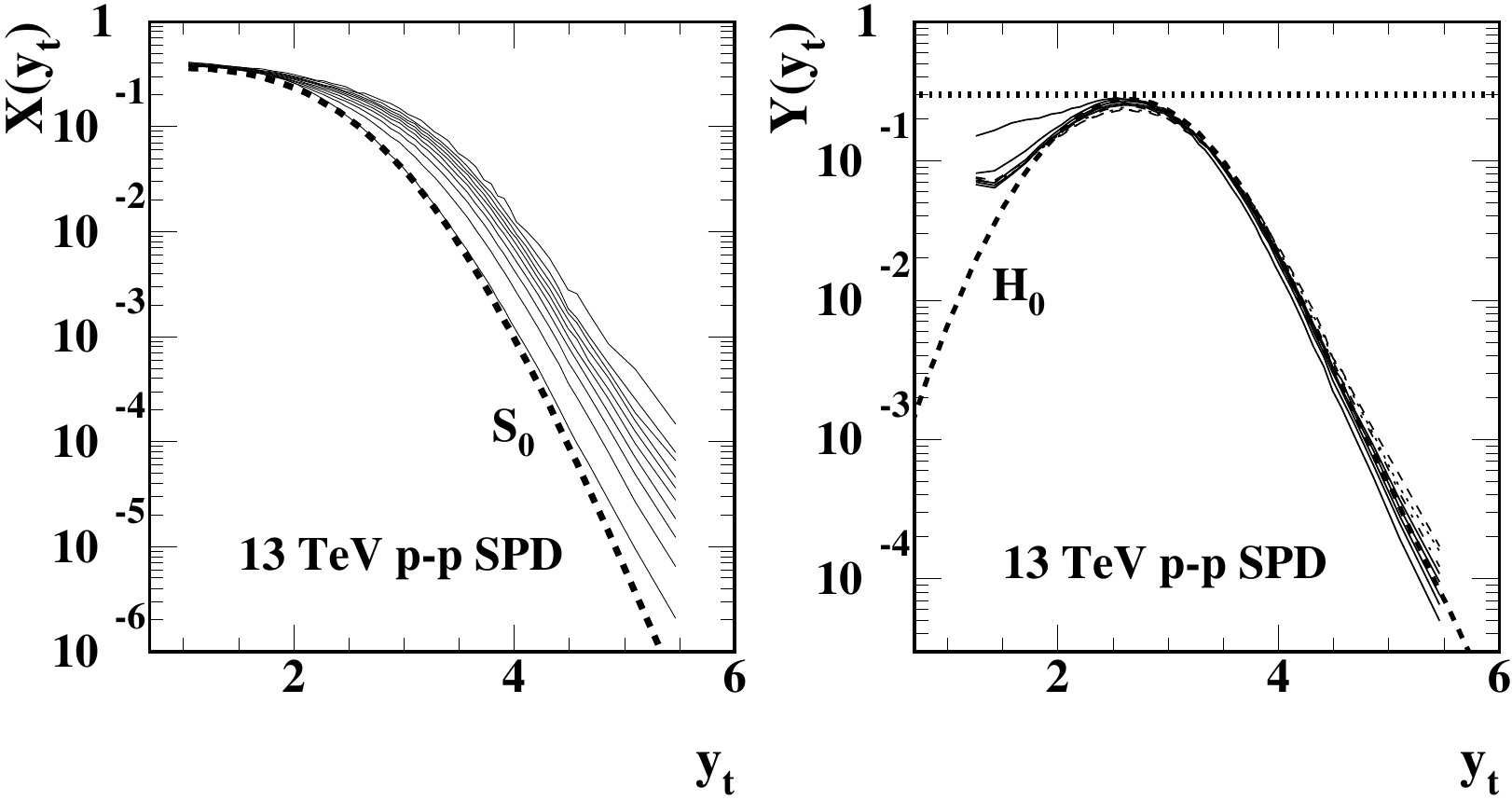}
	\put(-140,89) {\bf (c)}
	\put(-24,89) {\bf (d)}
	\caption{\label{tcm13}
		Left: Normalized \yt\ spectra in the form $X(y_t)$ from ten multiplicity classes of 13 TeV \pp\ collisions for V0M (a) and SPD (c) event selection.
		Right: Normalized spectrum hard components in the form $Y(y_t)$ for data in the left panels for  V0M (b) and SPD (d) event selection.  The bold dashed curves are TCM model functions.
	} 
\end{figure}

Figures~1 and 2 (a,c) of Ref.~\cite{newpptcm} show the general relation between the TCM (solid) and ALICE data (points). 
The TCM format of Figs.~\ref{tcm5} and $\ref{tcm13}$ below then provides a more-differential decomposition of spectrum data into soft and hard components. Panels (a,c) show full data spectra (thin solid) in the normalized form $X(y_t)$ defined above that are directly comparable with soft-component model $\hat S_0(y_t)$ (bold dashed). Below 0.5 GeV/c ($y_t \approx 2$) the data curves closely follow the model.  The same $\hat S_0(y_t)$ model is used for both event-selection methods.

Panels (b,d) show inferred data hard components $Y(y_t)$ defined by Eq.~(\ref{yyt}) (thin, several line styles) compared to TCM hard-component model $\hat H_0(y_t)$ (bold dashed). Deviations from $\hat H_0(y_t)$ below $y_t = 2$ appear in every \pp\ collision system (e.g.\ 200 GeV as reported in Ref.~\cite{ppprd}). The horizontal dotted lines provide a check on proper normalization of hard-component model $\hat H_0(y_t)$. The data hard component for the lowest multiplicity class is not shown because there is in effect very little jet contribution to those events due to strong selection bias. Note that full spectra in panels (a,c) for the lowest \nch\ class approximately coincide with $\hat S_0(y_t)$. The same $\hat H_0(y_t)$ model is used for both event-selection methods.

\subsection{Spectrum TCM parameter summary} \label{parsum}

Table~\ref{engparamy} presents TCM parameters for 5 and 13 TeV \pp\ collisions. Entries are grouped as soft-component parameters $(T,n)$, hard-component parameters $(\bar y_t,\sigma_{y_t},q)$, hard/soft ratio parameter $\alpha$ and NSD (non-single-diffractive) soft density $\bar \rho_{sNSD}$. For unidentified hadrons soft component $\hat S_0(m_t)$ may be approximated by Eq.~(\ref{s00}) for pions only, with parameters as in Table~\ref{engparamy}. Slope parameter $T = 145$ MeV is held fixed for all cases consistent with spectrum data.  Its value is determined  within a low-\yt\ interval where the hard component is negligible. 

\begin{table}[h]
	\caption{\pt\ spectrum TCM parameters for 5 TeV and 13 TeV NSD \pp\ collisions within $\Delta \eta \approx 2$.
	}
	\label{engparamy}
	\begin{center}
		\begin{tabular}{|c|c|c|c|c|c|c|c|c|} \hline
			$\sqrt{s}$ (TeV) & T\. (MeV) & $n$ & $\bar y_t$ & $\sigma_{y_t}$ & $q$ & $100\alpha$ & $\bar \rho_{s\text{NSD}}$ &  $\bar \rho_{0\text{NSD}}$ \\ \hline
			5.0 & 145  & 8.5 & 2.63 & 0.58  &  4.0  & 1.45 & 5.0 & 5.3 \\ \hline
			13.0  & 145  & 7.8 & 2.66 & 0.60  & 3.8  & 1.70 & 5.8 & 6.4 \\ \hline
		\end{tabular}
	\end{center}
\end{table}

For the present analysis  Eq.~(\ref{s00}) was evaluated separately for pions, charged kaons and protons with $T_i = 140$, 200 and 210 MeV respectively. Those expressions were then combined to form $\hat S_0(m_t)$ via Eq.~(\ref{s0m}) with $z_{0i} = 0.82$, 0.12 and 0.06 respectively. For each energy the same exponent $n$ was applied to three hadron species. L\'evy exponent $n$ values and hard-component $q$ values are as reported in Table~\ref{engparamy}. 
$\bar \rho_{sNSD}$ values are derived from the universal NSD trend $\bar \rho_{sNSD} \approx 0.81 \ln(\sqrt{s} / \text{10 GeV})$ inferred from spectrum and yield data. $\bar \rho_{0NSD}$ values are derived from the TCM relation $\bar \rho_0 \approx \bar \rho_s + \alpha \bar \rho_s^2$.

\section{\titlepp\ \titlept\ spectrum variable TCM} \label{tcmfit}

In a previous study of LHC \pp\ \pt\ spectra reported in Ref.~\cite{newpptcm} the TCM was used as a fixed reference independent of multiplicity \nch\ event class or event selection method (V0M or SPD). Deviations from the TCM revealed selection biases depending strongly on the event selection method. In the present study TCM parameters are varied to best accommodate data. The resulting parameter variations with charge multiplicity \nch\ and event selection (V0M vs SPD) provide information on physical mechanisms affected by selection methods. Emphasis is placed on 13 TeV data since 5 TeV trends are similar.

\subsection{Hard component $\bf \hat H_0(y_t;n_{ch})$ parameters}

Spectrum data indicate that no significant adjustment is required for spectrum TCM soft component $\hat S_0(y_t)$. Of the TCM hard-component $\hat H_0(y_t;n_{ch})$ parameters, $\bar y_t$ is the Gaussian centroid (mode), $\sigma_+$ is the Gaussian width {\em above} the mode, $q$ is the parameter controlling the exponential (power-law) tail and $\upsilon$ (upsilon) is a parameter controlling the width $\sigma_-$ {\em below} the mode according to
\bea
 \sigma_-^2 &\equiv& \sigma_+^2 / \upsilon.
\eea
$\hat H_0(y_t;n_{ch})$ is symmetric near the mode for $\upsilon = 1$. Reducing $\upsilon$ broadens the peak below the mode. That formulation is chosen to accommodate the requirement to represent very large widths below the mode for low \nch.

Parameter values for each \nch\ class and selection criterion are established by iterated visual comparison of data and model, not by automatic $\chi^2$ fits. Optimum parameter values are determined independently for each event class and are not readjusted thereafter. Each parameter influences a different \yt\ interval: $\sigma_-$ corresponds to $y_t < 2$, $\bar y_t$ to the difference slope near the mode, $\sigma_+$ to $y_t \in [2.7,4]$ and $q$ to $y_t > 4$. Thus it is relatively straightforward to optimize the data description.

\subsection{Spectrum ratios vs Z-scores}

Spectrum data-model comparisons have been conventionally represented by data/model ratios. Ratio values near 1 are interpreted to indicate acceptable models. However, that procedure can be quite misleading as discussed for instance in Ref.~\cite{ppprd}. A more effective measure of model validity is the Z-score~\cite{zscore} defined by
\bea \label{zscore}
Z_i &=& \frac{O_i - E_i}{\sigma_i} \rightarrow \frac{\text{data $-$ model}}{\text{error}},
\eea
where $O_i$ is a spectrum datum, $E_i$ is the corresponding expectation (model prediction) and $\sigma_i$ is the data r.m.s. statistical uncertainty (error). The r.m.s.\ average of Z-scores for an acceptable model should be near 1.
The relation between a data/model ratio and corresponding Z-scores is then given by
\bea \label{suppress}
\frac{\text{data}}{\text{model}} - 1 &\approx& \text{Z-scores} \times \frac{\text{error}}{\text{data}},
\eea
with error/model (exact) $\rightarrow $ error/data (approximate). Given the Z-score definition in Eq.~(\ref{zscore}) a variant of the $\chi^2_\nu$ statistic can be defined as
\bea \label{chinu}
\chi^2_\nu & \equiv & \frac{1}{\nu} \sum_{i=1}^N \frac{(O_i - E_i)^2}{\sigma_i^2} \approx \frac{1}{N} \sum_{i=1}^N Z_i^2
\eea
for $N$ data points in a spectrum. The approximation is that instead of number of degrees of freedom $\nu = N - \text{number of fit parameters}$, $N$ appears in the denominator.   For Poisson-distributed data $\sigma_i \approx \sqrt{E_i}$ in Eq.~(\ref{zscore}), leading to the formula often encountered for the $\chi^2$ statistic. Since the second above is the mean-squared Z-score it should also be near 1 for an acceptable model.

\begin{figure}[h]
	\includegraphics[width=1.65in]{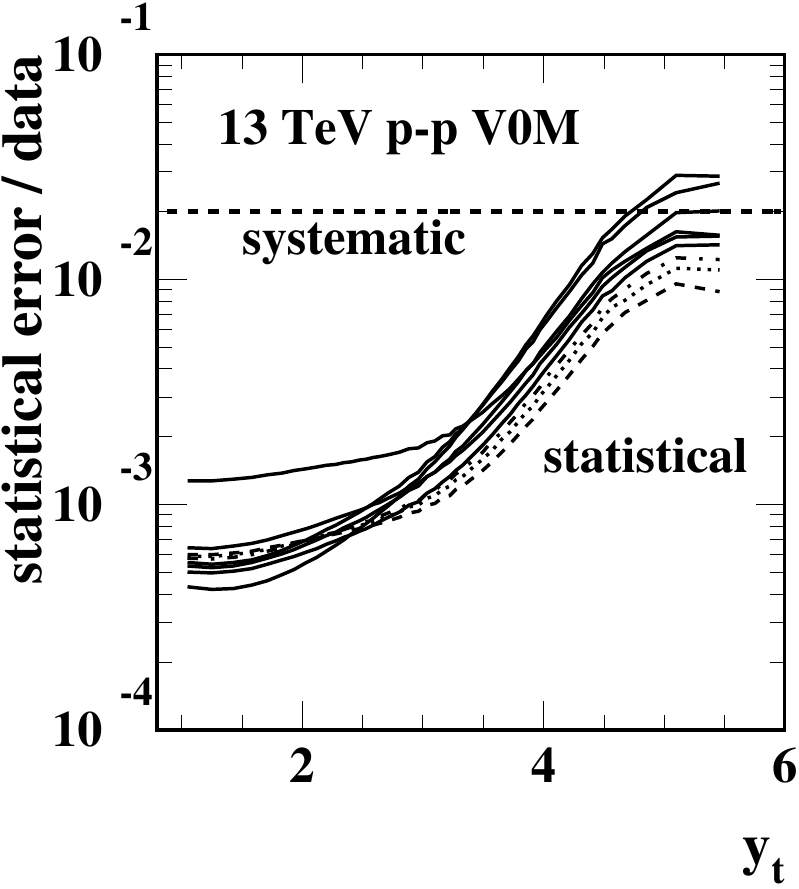}
	\includegraphics[width=1.68in]{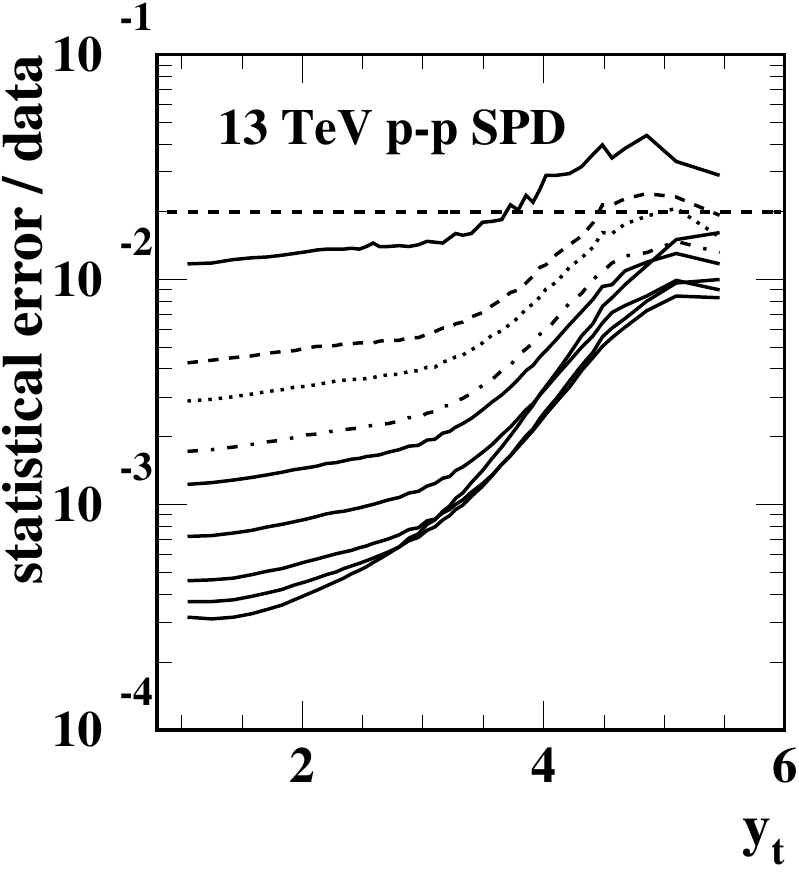}
	\caption{\label{errorrats}
	Published statistical uncertainties (errors) divided by corresponding data values for 13 TeV \pp\ collisions and for V0M (left) and SPD (right) event selection. The dashed line at left is explained in Sec.~\ref{sysdistort}.
	} 
\end{figure}

Figure~\ref{errorrats} shows error/data ratios for ten \nch\ classes of 13 TeV \pp\ spectra and for V0M and SPD event selection methods. Bunching (V0M) or spreading (SPD) of different event classes for the two methods results from the distribution of event {\em number} arising from the selection method. The distribution of magnitudes at lower \yt\ (dominated by the nonjet soft component) depends on event number. The relative magnitudes at higher \yt\ (jet-related hard components) also depend on the hard/soft ratio. It is interesting that the highest-\nch\ spectra (varying line styles) are lowest at high \yt\ in the left panel (V0M) but highest in the right panel (SPD). The top curve (solid) in the right panel corresponds to a relatively small number of $\bar \rho_0 \approx 54$ SPD events with large fractional jet contribution, about one third of total particles. 

The central message of this figure is that large Z-scores ($\gg 1$) that might falsify a model may be suppressed by orders of magnitude to produce misleading small deviations from 1 in spectrum {\em ratios}. Further, error/data ratios tend to vary on \yt\ over nearly two orders of magnitude in such a way that spectrum ratios might be visibly sensitive to deviations at higher \yt\ but completely insensitive to deviations at lower \yt\ {\em where the great majority of jet fragments resides}. Thus, models that may actually be dramatically falsified by data could appear {\em in ratios} to describe the data adequately It is useful to note that for an acceptable model [Z-scores $\approx O(1)$] deviations of spectrum ratios from 1 should approximate error/data ratios. In  the usual ratio plot format and for the present data statistics (see Fig.~\ref{errorrats}) such deviations would not be visible. Given  that preamble, TCM parametrizations for V0M and SPD event selection are presented in turn.

\subsection{TCM for V0M data}

\begin{figure}[h]
	\includegraphics[width=3.3in]{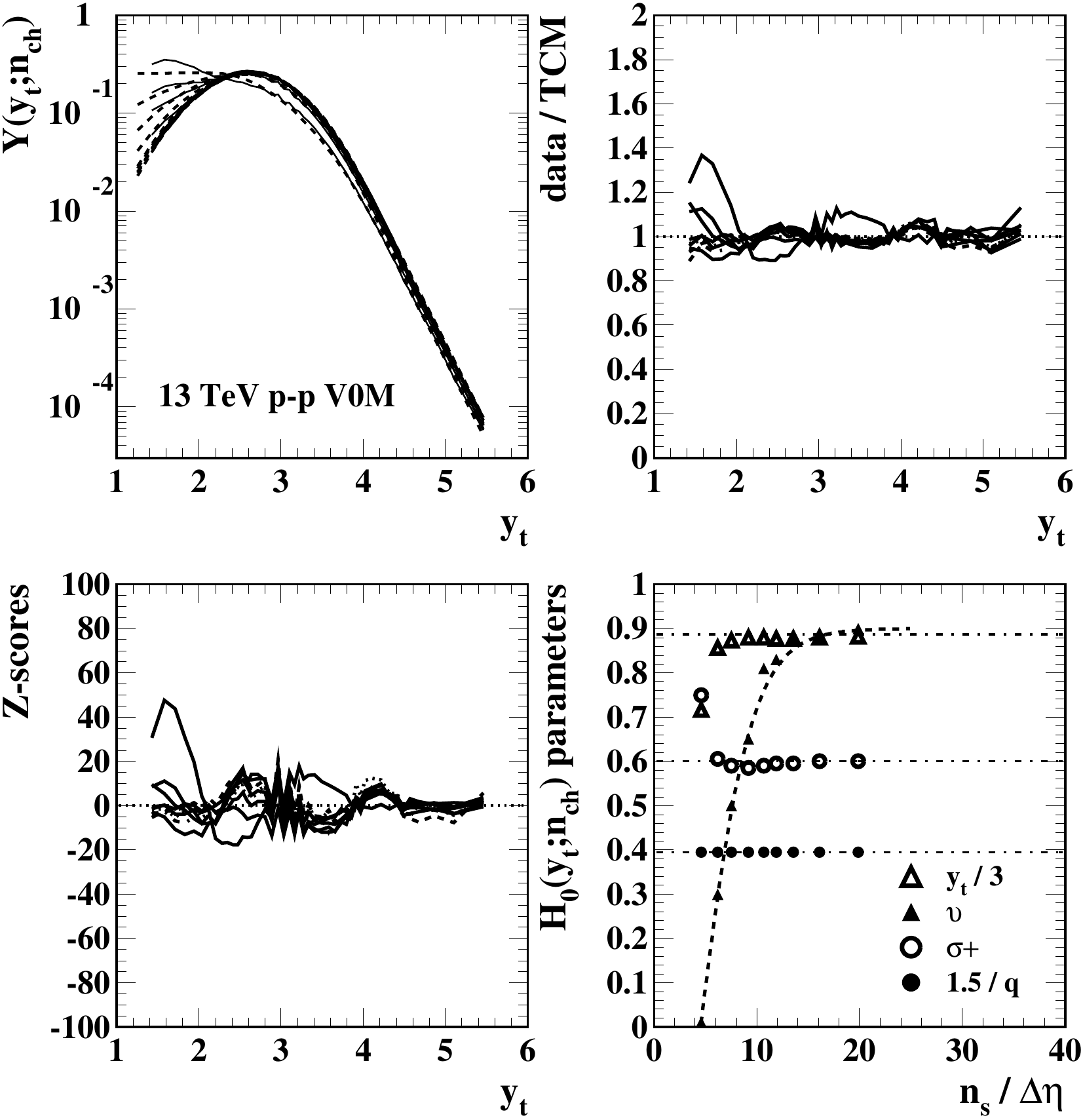}
	\put(-150,220) {\bf (a)}
	\put(-24,220) {\bf (b)}
	\put(-150,90) {\bf (c)}
	\put(-24,90) {\bf (d)}
	\caption{\label{v0mfit}
		(a) Spectrum data hard components in the form $Y(y_t)$ (solid) compared to TCM variable hard-component model $\hat H_0(y_t;n_{ch})$ (dashed) for nine event classes of 13 TeV \pp\ collisions and V0M event selection.
		(b) Corresponding data/model ratios.
		(c) Corresponding Z-scores.
		(d) $\hat H_0(y_t;n_{ch})$ parameter variations with \nch.
	} 
\end{figure}
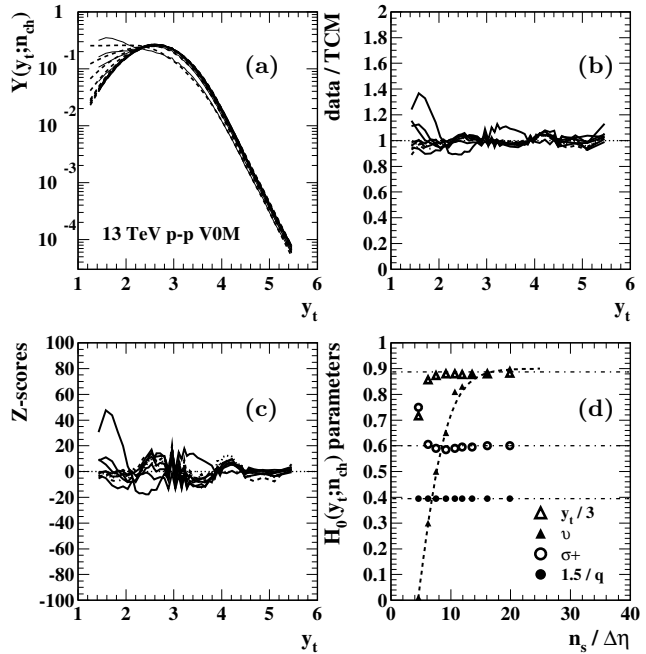

Figure~\ref{v0mfit} (a) shows data hard components (solid) from \pt\ spectra corresponding to V0M event selection applied to 13 TeV \pp\ data. Those spectra correspond to Fig.~\ref{tcm13} (b) of the present article with quantity $Y(y_t;n_{ch})$ as defined in Eq.~(\ref{yyt}) and directly comparable with TCM hard component $\hat H_0(y_t;n_{ch})$. The dashed curves show varying model $\hat H_0(y_t;n_{ch})$ with parameters adjusted to best accommodate the data as described below.

Figure~\ref{v0mfit} (b) shows data/TCM hard-component ratios. Whereas the plots of full spectra in Fig.~\ref{tcm13} (b) include ten event classes the plots in this figure exclude class 10 corresponding to the lowest \nch\ value  $\bar \rho_0 = 2.54$ where the jet contribution is strongly suppressed. Most deviations from 1 are less than 10\% suggesting good agreement between data and model. The event class with the largest deviations is class 9 with the lowest charge density $\bar \rho_0 = 4.94$, substantially below the NSD value 6.4.

Figure~\ref{v0mfit} (c) shows Z-scores as defined by Eq.~(\ref{zscore}). Whereas the data/TCM ratios in panel (b) suggest an acceptable model description the actual Z-scores are $O(10) \gg 1$ and thus {\em nominally} justify model rejection. However, two characteristics should be noted: (a) The Z-score trends are very consistent for all but the lowest plotted \nch\ class ($n = 9$). (b) The characteristic length scale of significant variations for $n \in [1,8]$ is small, inconsistent with the basic physics of both the jet-related data hard component and the TCM hard-component model. That issue is especially obvious for the oscillatory structure near $y_t = 3$. Those properties suggest that the residual structure may be due to artifacts imposed on the spectra, possibly during inefficiency correction or as a result of local \pt\ calibration issues.

Figure~\ref{v0mfit} (d) shows resulting parameter trends for 13 TeV $\hat H_0(y_t;n_{ch})$. For varying $\hat H_0(y_t;n_{ch})$, the V0M parameter trends as \nch\ {\em decreases} are as follows: (a) For several high-multiplicity classes the parameter values are consistent with the fixed model adopted in Ref.~\cite{newpptcm} (dash-dotted lines). (b) Parameter $\upsilon$ decreases rapidly toward zero to accommodate strong width $\sigma_{y_t-}$ increases below the mode in panel (a). (c) For  the lowest two \nch\ classes the combination of mode $\bar y_t$ (decreasing), upper width $\sigma_{y_t+}$ (increasing) and lower width $\sigma_{y_t-}$ (increasing dramatically) effectively moves the peak mode down on \yt\ while the exponential tail {\em remains nearly stationary}. (d) The data do not require variation of tail parameter $q$ (or equivalently, power-law exponent $n$), consistent with observations based on logarithmic derivatives in Ref.~\cite{newpptcm}.

The 13 TeV trend for $\upsilon$ (dashed) has the same $\tanh$ form as the 200 GeV trend reported in Ref.~\cite{alicetomspec} Fig.~6 (left). The maximum value of $\upsilon$ in this case is 0.9 whereas it is near 3 for 200 GeV \pp\ data. Low-\yt\ width $\sigma_{y_t-}$ is always greater than $\sigma_{y_t+}$ for 13 TeV, whereas $\sigma_{y_t-}$ falls well below $\sigma_{y_t+}$ for 200 GeV spectra and larger \nch.

\begin{figure}[h]
	\includegraphics[width=3.3in]{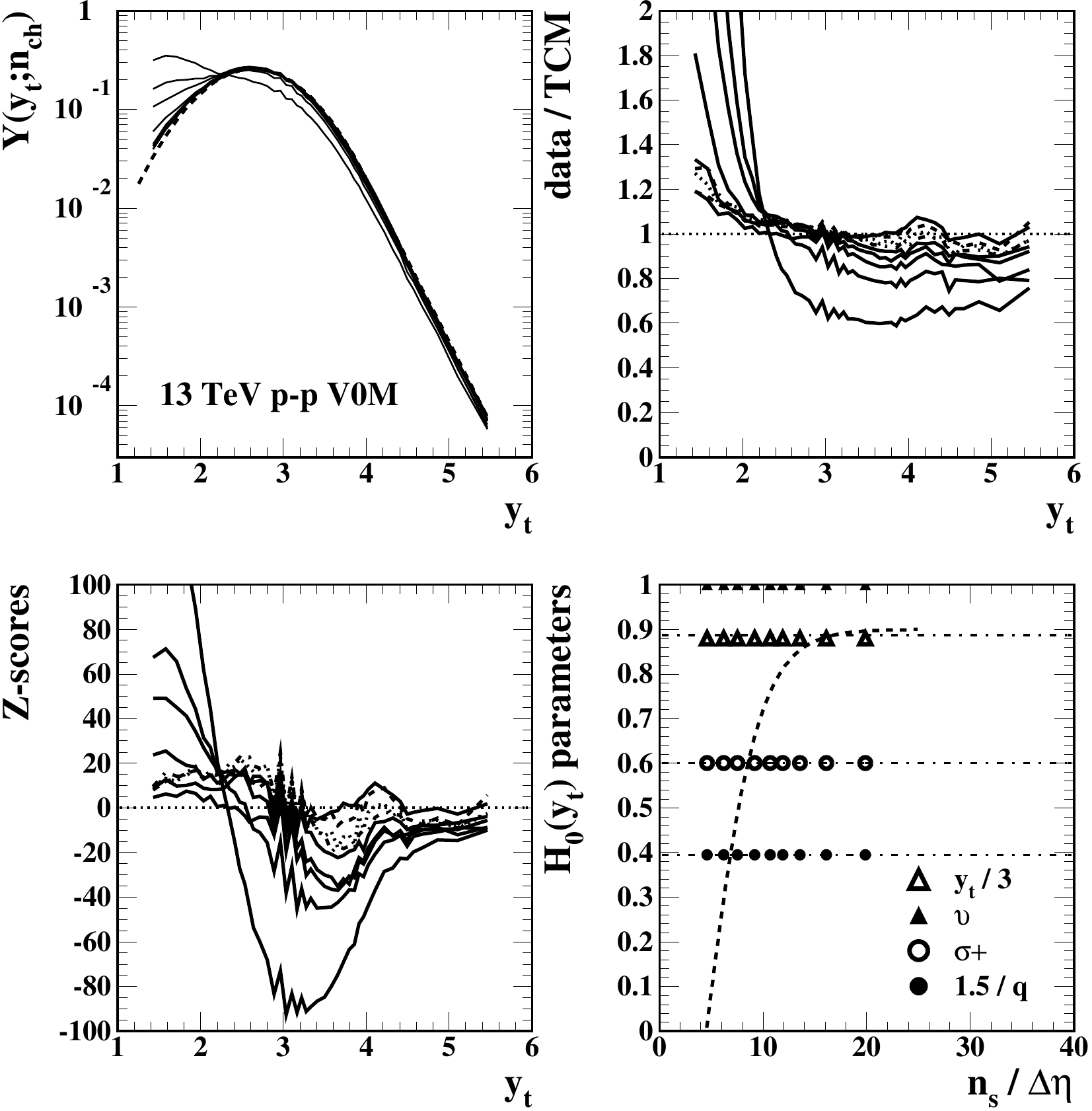}
\put(-150,220) {\bf (a)}
\put(-24,220) {\bf (b)}
\put(-150,90) {\bf (c)}
\put(-24,90) {\bf (d)}
\caption{\label{v0mnofit}
		(a) Spectrum data hard components in the form $Y(y_t)$ (solid) compared to TCM {\em fixed} hard-component model $\hat H_0(y_t)$ (dashed) for nine event classes of 13 TeV \pp\ collisions and V0M event selection.
(b) Corresponding data/model ratios.
(c) Corresponding Z-scores.
(d) $\hat H_0(y_t)$ fixed parameters.
	} 
\end{figure}
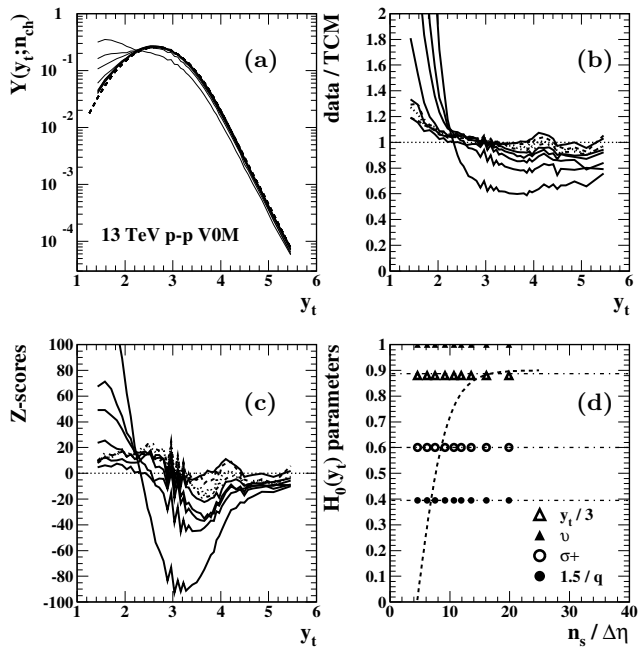

Fig.~\ref{v0mnofit} shows results in the same plot formats if the TCM is restricted to fixed model functions as in Ref.~\cite{newpptcm}. Panel (a) is then equivalent to Fig.~\ref{tcm13} (b) of the present study. Panel (b) corresponds to Fig.~2 (b) of Ref.~\cite{newpptcm} (full spectrum ratios) with the following differences: (a) The lowest \nch\ class ($n = 10$) is not shown for the hard component alone. (b) Since the soft component is missing from both data and model in the ratio the present result should differ increasingly from Ref.~\cite{newpptcm} for lower \yt\ values.

Fig.~\ref{v0mnofit} (c) shows corresponding Z-scores that can be compared with Fig.~8 (b) of Ref.~\cite{newpptcm}. Again, the $n = 10$ data are omitted. The y-axis scale is chosen as a compromise between adequate sensitivity to observe detailed structure and comparisons with results for the fixed TCM model with much larger Z-score values. Panel (d) confirms that the $\hat H_0(y_t;n_{ch})$ parameters are held fixed.

\subsection{TCM for SPD data}

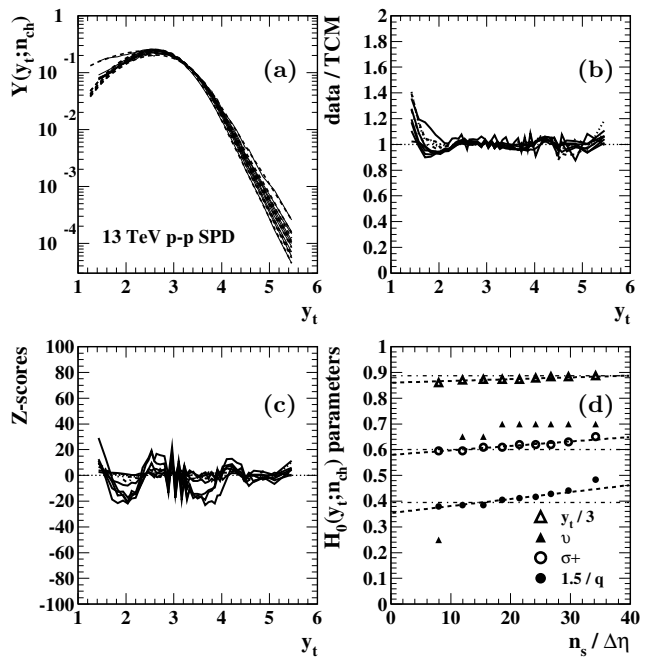
\begin{figure}[h]
	\includegraphics[width=3.3in]{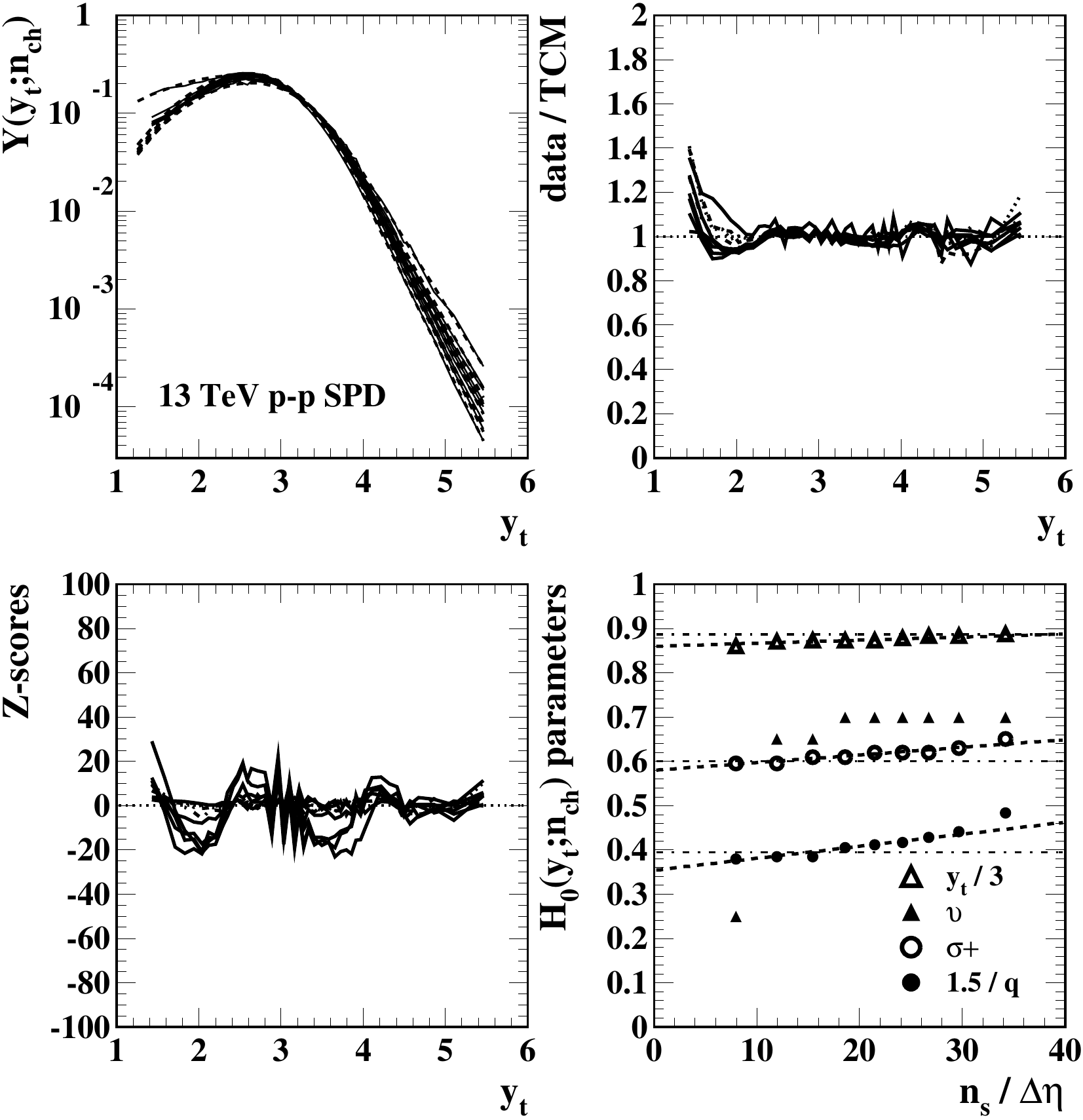}
\put(-143,220) {\bf (a)}
\put(-24,220) {\bf (b)}
\put(-143,93) {\bf (c)}
\put(-24,93) {\bf (d)}
\caption{\label{spdfit}
		(a) Spectrum data hard components in the form $Y(y_t)$ (solid) compared to TCM variable hard-component model $\hat H_0(y_t;n_{ch})$ (dashed) for nine event classes of 13 TeV \pp\ collisions and SPD event selection.
(b) Corresponding data/model ratios.
(c) Corresponding Z-scores.
(d) $\hat H_0(y_t;n_{ch})$ parameter variations with \nch.
	}  
\end{figure}

Fig.~\ref{spdfit} (a) shows spectrum hard components for SPD event selection equivalent to Fig.~\ref{tcm13} (d) of the present article. Data/model spectrum ratios (b) are similar to those for V0M data. Unlike the V0M Z-score trend SPD Z-scores (c) vary strongly in amplitude depending on \nch\ class. The strong variation results from the large spread in error/data ratio values in Fig.~\ref{errorrats} (right). For higher-multiplicity SPD data error/data ratios are large and the Z-score amplitudes are thus near 1 -- model and data are then {\em nominally} statistically consistent. Event classes with smaller \nch\ exhibit substantially larger (factor 10 or more) Z-score amplitudes because error/data ratios are ten to fifty times smaller than for higher-\nch\ data.

Fig.~\ref{spdfit} (d) shows $\hat H_0(y_t;n_{ch})$ parameter trends vs $\bar \rho_s$ for SPD event selection. The SPD parameter trends are as follows: (a) Centroid $\bar y_t$ and width above the mode $\sigma_{y_t+}$ increase, shifting the mode {\em and} the exponential tail to higher \yt. Width below the mode $\sigma_{y_t-}$ is generally greater than for V0M ($\upsilon$ smaller) but increases significantly only for the lowest \nch\ class. Exponential-tail parameter $q$ decreases substantially with increasing \nch\ ($1.5/q$ increases substantially) as noted in Ref.~\cite{newpptcm} in relation to a log-derivative method applied to spectra. 

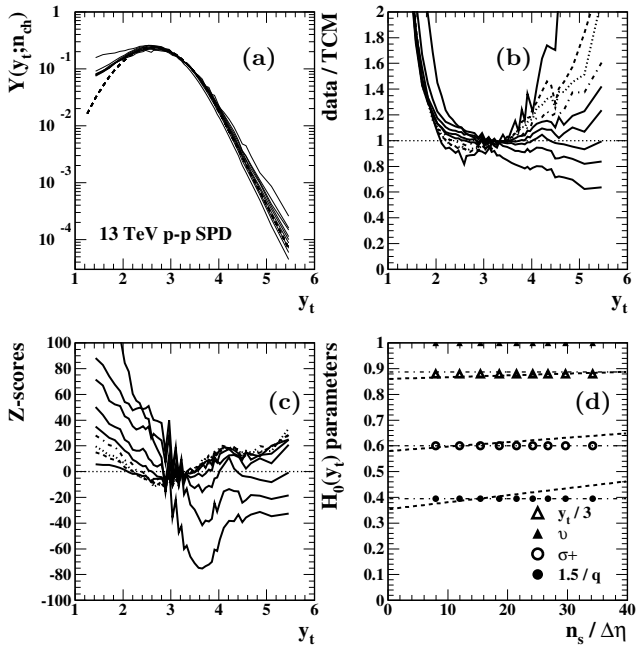
\begin{figure}[h]
	\includegraphics[width=3.3in]{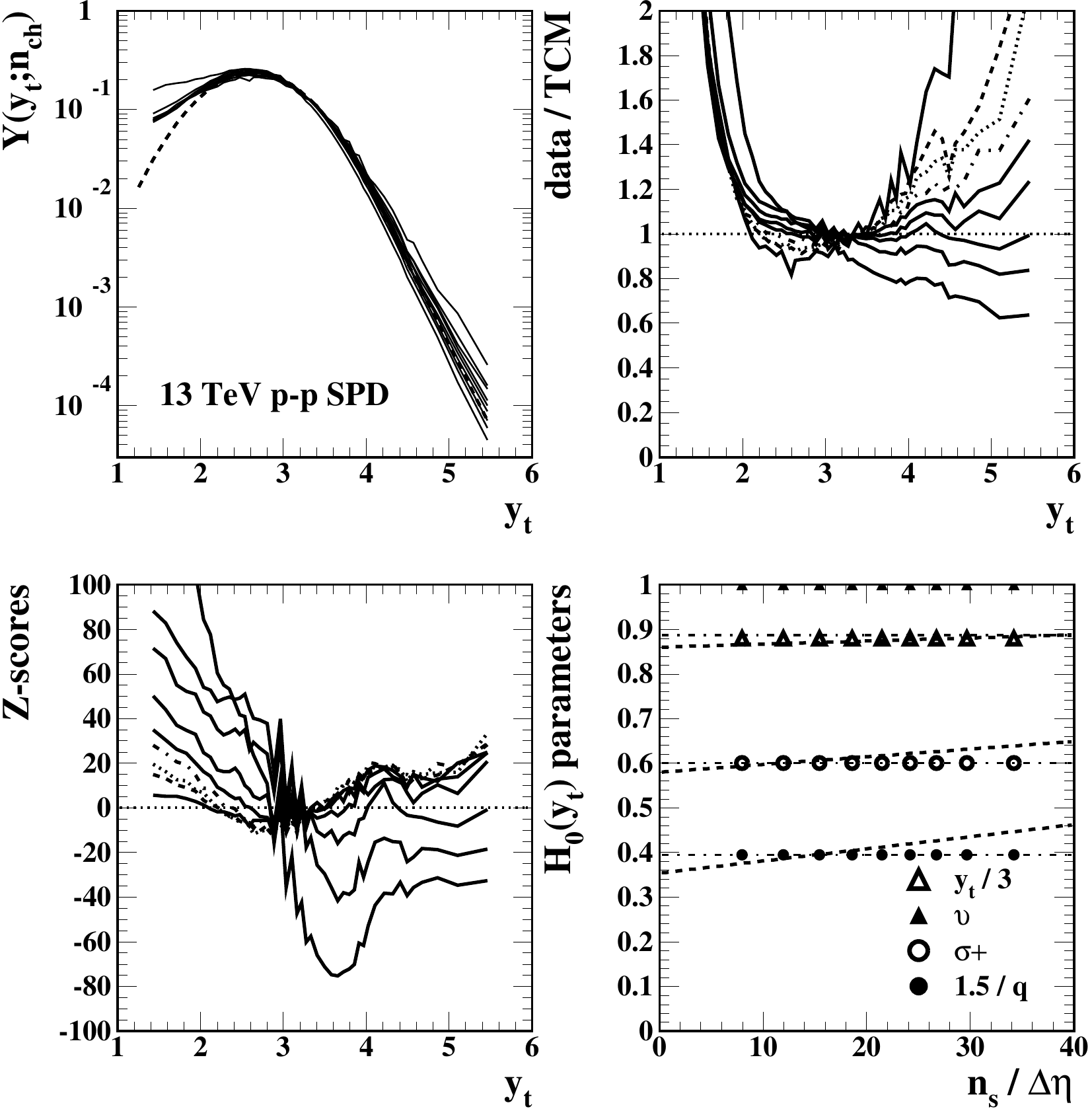}
\put(-150,220) {\bf (a)}
\put(-54,220) {\bf (b)}
\put(-140,90) {\bf (c)}
\put(-24,90) {\bf (d)}
\caption{\label{spdnofit}
		(a) Spectrum data hard components in the form $Y(y_t)$ (solid) compared to TCM {\em fixed} hard-component model $\hat H_0(y_t)$ (dashed) for nine event classes of 13 TeV \pp\ collisions and SPD event selection.
(b) Corresponding data/model ratios.
(c) Corresponding Z-scores.
(d) $\hat H_0(y_t)$ fixed parameters.
	}  
\end{figure}

Fig.~\ref{spdnofit} shows corresponding results for the TCM held fixed as in Ref.~\cite{newpptcm}. Panel (b) corresponds to Fig.~2 (d) of Ref.~\cite{newpptcm} with the same caveats as for Fig.~\ref{v0mnofit} above.  Panel (c) corresponds to Fig.~8 (d) of Ref.~\cite{newpptcm}, again with the same caveats. Panel (d) confirms the fixed model. Figures~\ref{v0mfit} (d) and \ref{spdfit} (d) demonstrate that the fixed TCM of Ref.~\cite{newpptcm} provides a reasonable compromise among event-selection methods and varying $\hat H_0(y_t;n_{ch})$ parameters.

\subsection{Summary}

In terms of $\hat H_0(y_t;n_{ch})$ parameter trends V0M and SPD event-selection methods are complementary. For V0M the width below the mode $\sigma_{y_t-}$ varies strongly and nonlinearly whereas structure above the mode remains fixed for all but the two lowest \nch\ classes (where the mode is shifted down on \yt). For SPD the structure above the mode varies strongly {\em and linearly} (with $\bar \rho_s$) whereas  $\sigma_{y_t-}$ (and $\upsilon$) remains nearly fixed except for the lowest \nch\ class.
For 13 TeV and large \nch\ $\sigma_{y_t-}$ for V0M decreases to just above the NSD width 0.6 while the SPD width remains substantially above that value. In contrast, for 200 GeV data (effectively SPD) $\sigma_{y_t-}$ decreases to 2/3 of the NSD width ($\upsilon$ increases to $\approx 3$ at 200 GeV whereas it remains below 1 for 13 TeV).

If model and data are statistically compatible the resulting Z-scores [$\approx O(1)$] should be relatively independent of event number and therefore of statistical error. That implies deviations in the form data/model - 1 should {\em vary} proportional to error/data ratios. However, if  there is a significant data - model deviation, either due to incorrect model or to systematic data imperfections, then data/model ratios should remain relatively independent of statistical error (or event number) whereas Z-scores should vary proportional to data/error. An example is found in the comparison of Fig.~\ref{v0mfit} (c) and Fig.~\ref{spdfit} (c). In the former case the Z-score trends, although exhibiting substantial \yt\ variation, are tightly grouped in overall amplitude corresponding to Fig.~\ref{errorrats} (left), whereas in the latter case although the {\em shapes} of the Z-score distributions are similar to V0M shapes the {\em amplitudes} vary over more than an order of magnitude consistent with Fig.~\ref{errorrats} (right). In contrast, panels (b) are similar. 

Given the statistical similarity of deviations in panels (b) it may be reasonable to introduce systematic uncertainty estimates $\approx 0.02$ (see Fig.~\ref{rms}, right) that would combine with statistical errors in quadrature to determine $\sigma_i$ in Eq.~\ref{zscore}. In effect, a {\em fractional} systematic uncertainty of 0.02 would establish a lower bound (dashed lines) on error/data ratios in Fig.~\ref{errorrats}.

The different systematic variations of the data hard component with \nch\ for V0M and SPD event selection can be interpreted in terms of how each selection method responds to strong fluctuations in the shape of the hard component above and below its mode. In either case the {\em same event population and jet population} are subject to selection. For V0M selection fluctuations above the mode are averaged the same for all \nch\ classes but V0M couples strongly to fluctuations below the mode. For SPD selection the complementary situation prevails.

In terms of physics interpretation the parameter trends demonstrate that variations at higher \yt\ (where jets dominate a spectrum) are closely correlated with variations at lower \yt\ (where nonjet processes {\em numerically} dominate the {\em full} spectrum), strongly supporting the interpretation that what emerges as the {\em hard} component of TCM analysis represents the complete jet contribution to \pp\ spectra, with maximum contribution near $p_t = 1$ GeV/c. The V0M and SPD differences in $\hat H_0(y_t;n_{ch})$  parameter trends suggest that different $\eta$ acceptances are sensitive to distinct elements of the overall chain from proton internal structure (PDF) to midrapidity low-$x$ gluon jets.

\section{Tsallis spectrum model} \label{tsallismodel}

In the context of recent conjectures that a QGP may appear in higher-multiplicity \pp\ collisions at LHC energies based on interpretation of certain data features~\cite{moreppflow,dusling,thoughts}, two spectrum models that relate to partial equilibration and flows within multiparticle systems have been invoked to describe \pp\ \pt\ spectrum data. As an example, Ref.~\cite{cleymans} has applied Tsallis and blast-wave (BW) models to \pp\ spectrum data from Ref.~\cite{alicenewspec}. Those results are reviewed in this section and the next.

The Tsallis (non-extensive statistics) fit model~\cite{tsallis,tsallis1} is applied to V0M \pt\ spectra for 5 and 13 TeV \pp\ collisions~\cite{cleymans}. It has been argued that in high multiplicity LHC \pp\ collisions multiple parton collisions could lead to thermalization. The Tsallis model and its parameter $q$ might then measure ``the departure of the system from an equilibrium state.'' The Tsallis model is said to provide ``a  very good description of experimental data for the complete transverse momentum range in various multiplicity classes and collision energies.'' But other statements in Ref.~\cite{cleymans} seem to contradict that assertion.

\subsection{Tsallis method}

Non-extensive statistics is parametrized in part by Tsallis (not TCM) parameter $q$. Difference $q - 1$ is interpreted as a measure of departure from equilibrium in a heterogeneous system~\cite{tsallis}. The Tsallis $q$-exponential probability distribution on $x$ is presented in Ref.~\cite{cleymans} as
\bea \label{tsalliseq}
\exp_q(- x/T) &\equiv& \frac{1}{[1 + (q-1) x/T ]^{1/(q-1)}}
\\ \nonumber
&=& \frac{1}{(1 + x/nT )^n},
\eea
where the second line arises from taking $q-1 \rightarrow 1/n$ and may be compared with Eq.~(\ref{s00}). Parameter $n$ is then the ``power'' in the so-called power-law model $\propto 1/x^n$. In the limit $1/n \rightarrow 0$ Eq.~(\ref{tsalliseq}) goes to exponential $ \exp(-x/T)$.

Although the spectrum model invoked in Ref.~\cite{cleymans} is nominally associated with Tsallis statistics the actual form used to fit spectrum data (for hadron species $i$) in Ref.~\cite{cleymans} is significantly different compared to the expression in Eq.~(\ref{tsalliseq}) (second line):
\bea \label{badtsallis}
\frac{d^2n_{ch,i}}{p_t dp_t d\eta} &=& w_i \bar \rho_0 \left[\frac{p_t C_i}{(1+m_{t,i}/nT)^{n+1}}\right],
\eea
i.e.\ the $q$-exponential has been raised to power $q$ (leading to $n+1$ as the exponent in the denominator) and there is an extra factor $p_t$. The motivating statement is ``...an additional power of $q$...is necessary to make the Tsallis statistics thermodynamically consistent.'' The extra factor \pt\  (for $dn_{ch}/d\eta$ or $m_t$ for $dn_{ch}/dy_z$) is not explained. However, for large \pt\ Eq.~(\ref{badtsallis}) goes asymptotically to $1/p_t^n$ which is then consistent with Eq.~(\ref{tsalliseq}) in that limit. 

\subsection{Tsallis results}

\begin{figure}[h]
	\includegraphics[width=3.3in,height=1.6in]{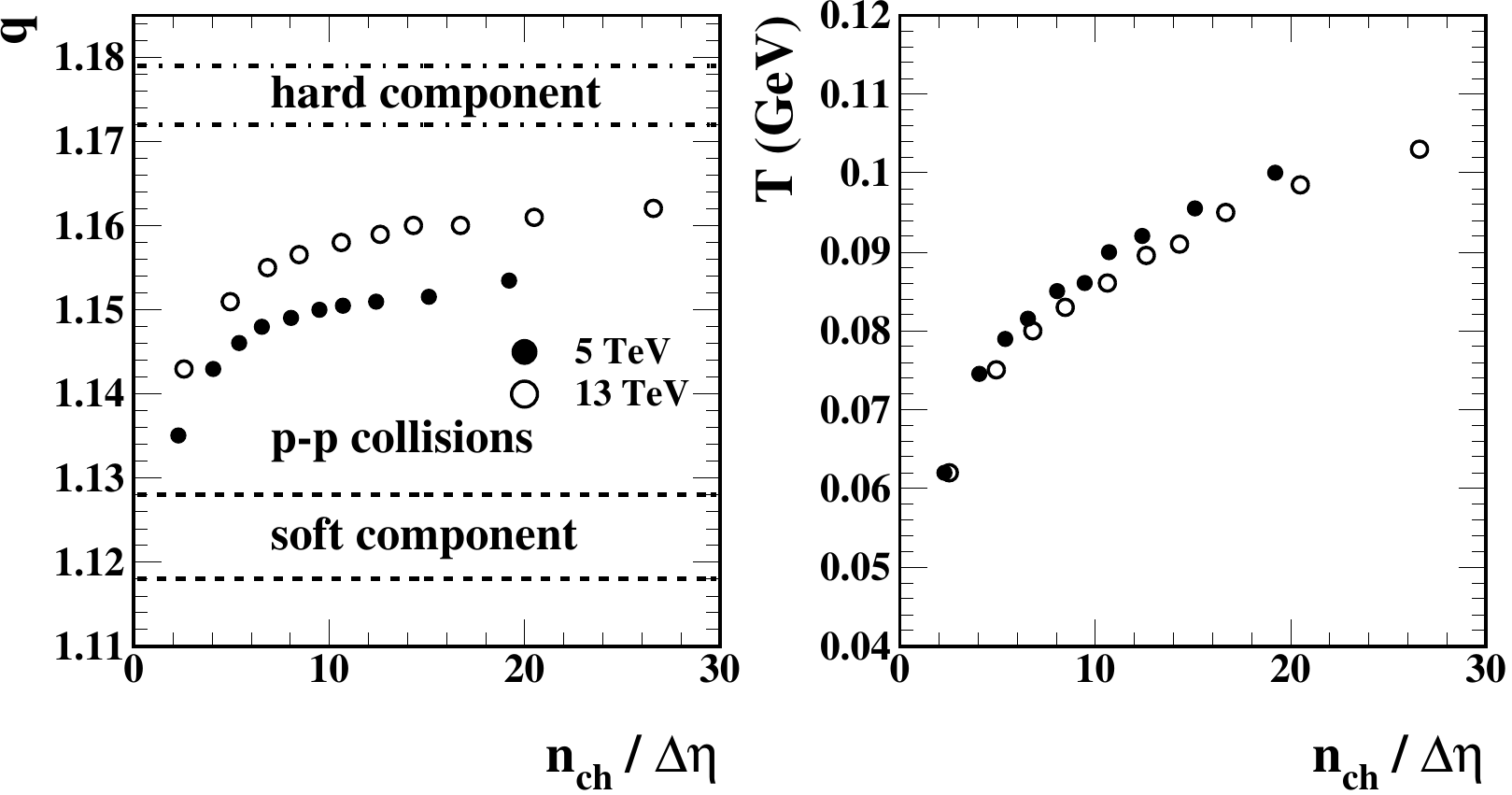}
	\caption{\label{qt}
		Tsallis-model fitted values for parameters $q$ (left) and $T$ (right) from ten \nch\ classes each of 5 TeV (solid dots) and 13 TeV (open circles) V0M spectra. Dashed and dash-dotted lines in the left panel refer to TCM limiting cases.
	} 
\end{figure}

Figure~\ref{qt} shows Tsallis parameters $q$ and $T$ derived from fits to \pt\ spectra from ten V0M multiplicity classes of 5 and 13 TeV \pp\ collisions from Ref.~\cite{alicenewspec}. The parameter values are taken from Figs.~3 and 4 of Ref.~\cite{cleymans}. Those parameter values are used in what follows. The constants $C_i$ in Eq.~(\ref{badtsallis}) are determined such that the expression in square brackets is unit normal for each hadron species. The weights $w_i$ are from Ref.~\cite{cleymans} and densities $\bar \rho_0$ are from Table~1 of Ref.~\cite{alicenewspec}. Models for three charged-hadron species $i$ with weights $w_i$ are summed to form the model for $\bar \rho_0(y_t)$ describing unidentified hadrons.

It is of interest to compare fitted Tsallis parameter values with values for the TCM. The TCM soft-component exponent is $n \approx 8.5$ and 7.8 for 5 and 13 TeV respectively. The corresponding Tsallis $q = 1 + 1/n$ values (dashed) are then 1.118 and 1.128 respectively. The ``power-law'' exponents $n$ in Fig.~9 (right) of Ref.~\cite{newpptcm} are 5.80 and 5.60 with corresponding Tsallis $q$ values (dash-dotted) 1.172 and 1.179. With increasing \nch\ Tsallis $q$ increases strongly above the TCM soft-component lower limits toward the hard-component values as hard/soft ratio $x \approx \alpha \bar \rho_s$ increases with \nch. Fitted values of $T$ should be compared with fixed TCM value 145 MeV for unidentified hadrons that is the same for any collision system.

\begin{figure}[h]
	\includegraphics[width=3.3in,height=1.6in]{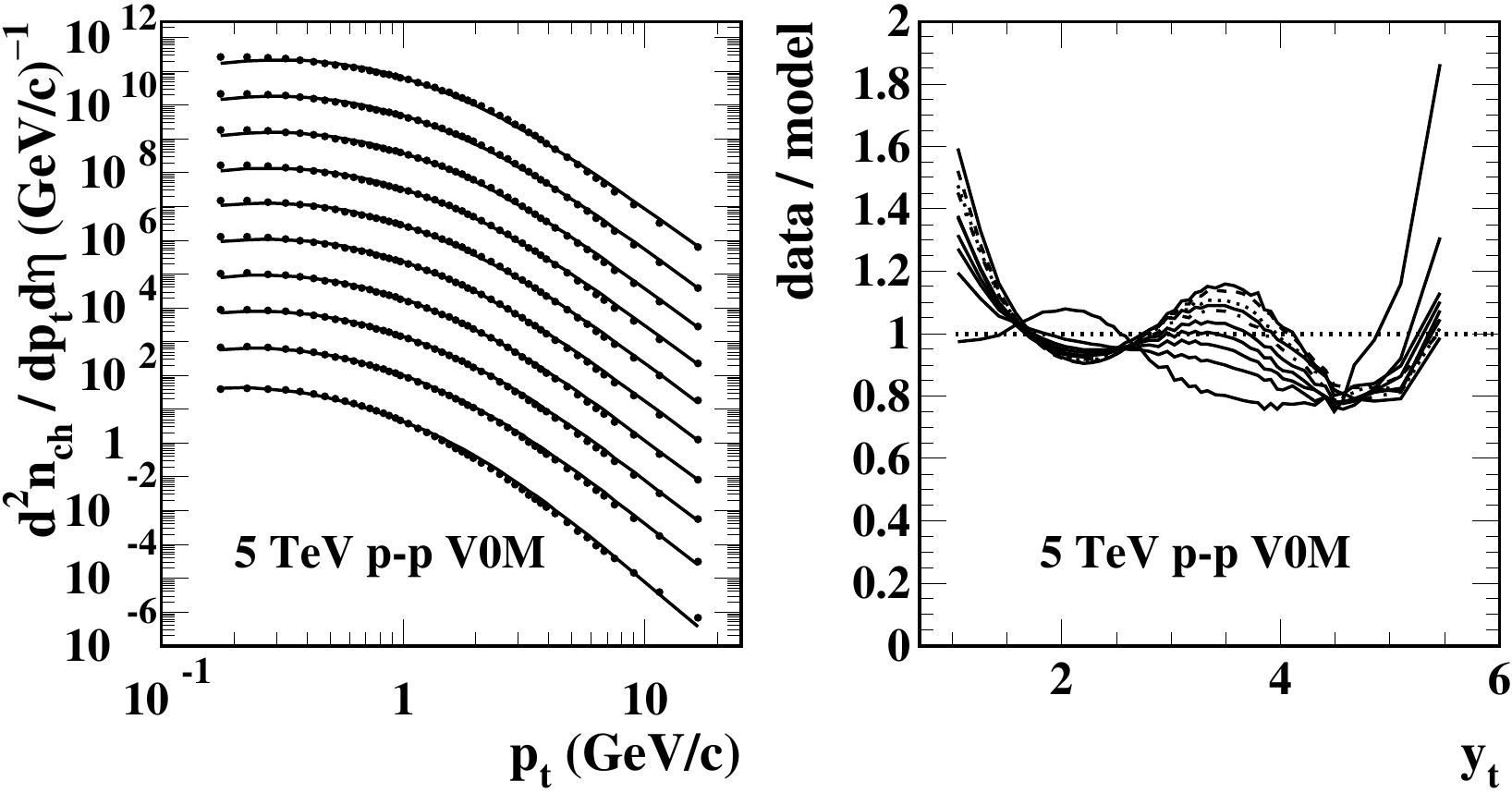}
\put(-140,99) {\bf (a)}
\put(-26,99) {\bf (b)}\\
	\includegraphics[width=3.3in,height=1.6in]{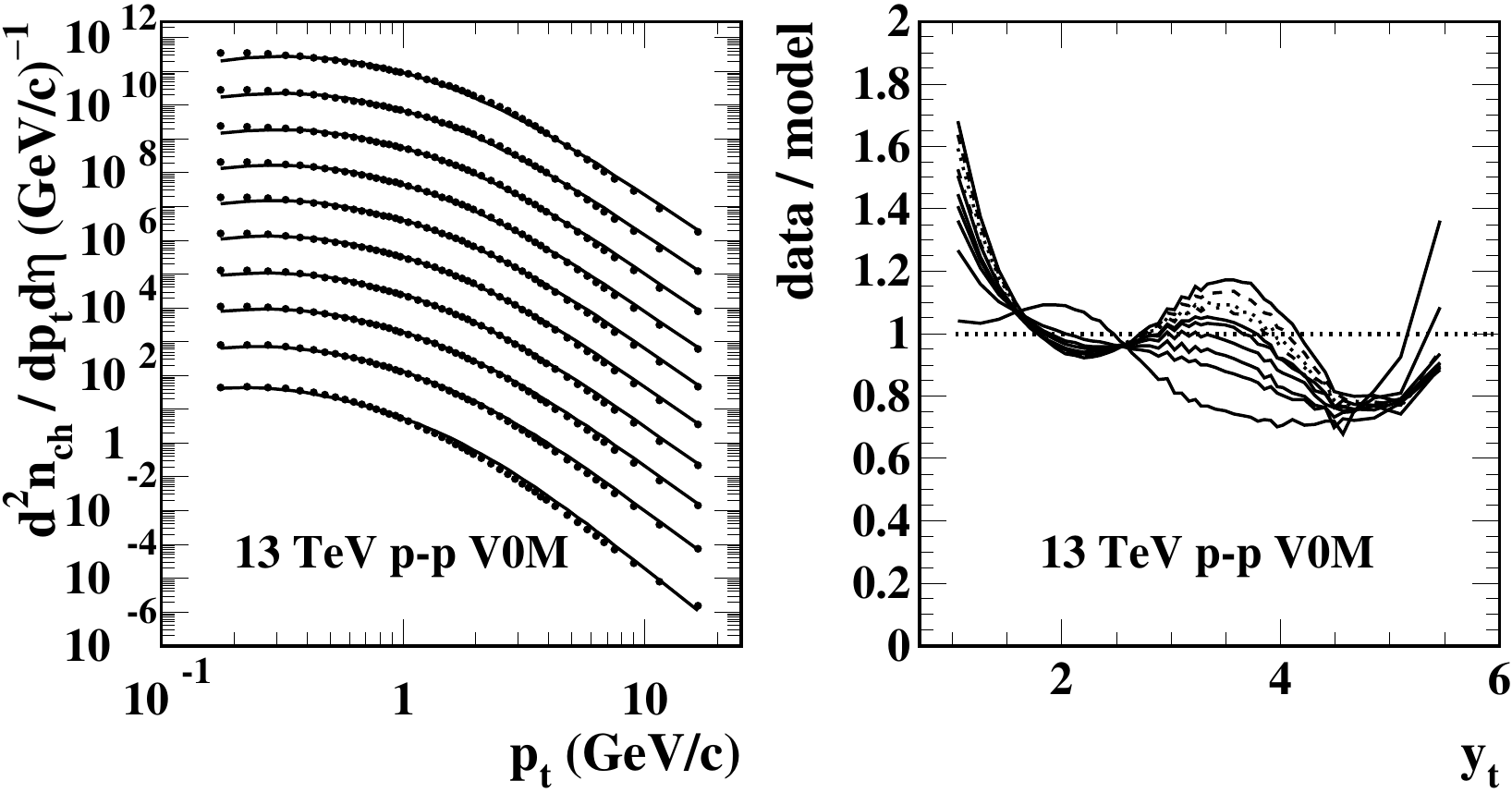}
\put(-140,99) {\bf (c)}
\put(-26,99) {\bf (d)}
	\caption{\label{tsallisfits}
		Left: Tsallis model fits (solid) to \pt\ spectra (points) from 5 TeV (a) and 13 TeV (c) \pp\ collisions and V0M event selection. The Tsallis spectrum model is as shown in Eq.~(\ref{badtsallis}) and as reported in Ref.~\cite{cleymans} with parameters from Fig.~\ref{qt}.
		Right: Data/model ratios for results in the left panels.
	} 
\end{figure}

Figure~\ref{tsallisfits} (a,c) shows the Tsallis model [solid curves described by Eq.~(\ref{badtsallis})] and 5 and 13 TeV \pp\ \pt\ spectrum data (points) from Ref.~\cite{alicenewspec} corresponding to Figs.~\ref{tcm5} (a) and \ref{tcm13} (a) in Sec.~\ref{tcmspecdat} based on V0M event selection. The Tsallis parameter values are those shown in Fig.~\ref{qt}. 

Figure~\ref{tsallisfits} (b,d) shows corresponding data/model ratios that are consistent with Figs.~1 and 2 of Ref.~\cite{cleymans}, although important details at lower \pt\ are more visually accessible on transverse rapidity \yt. Reference~\cite{cleymans} concludes ``From the ratio between the experimental data points and the fit function, it is observed that the non-extensive statistics [Tsallis model] provides a good description of the charged particle transverse momentum spectra for the complete $p_T$ region.'' It is notable that  spectrum ratios turn up sharply at lower \pt, which may be related to the extra factor $p_t$ in Eq.~(\ref{badtsallis}). And there is a qualitative difference in form of the fit to the lowest-multiplicity spectra. Although there are clearly substantial model-data deviations this ratio format conceals the statistical significance differentially as  a function of \pt.

\begin{figure}[h]
	\includegraphics[width=1.65in,height=1.6in]{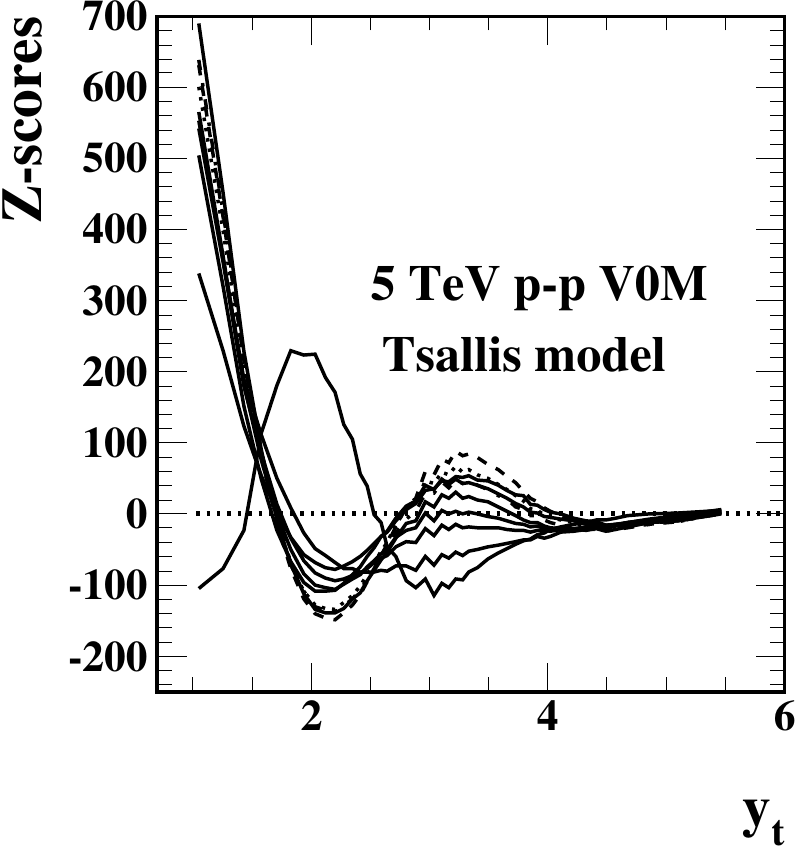}
	\includegraphics[width=1.65in,height=1.6in]{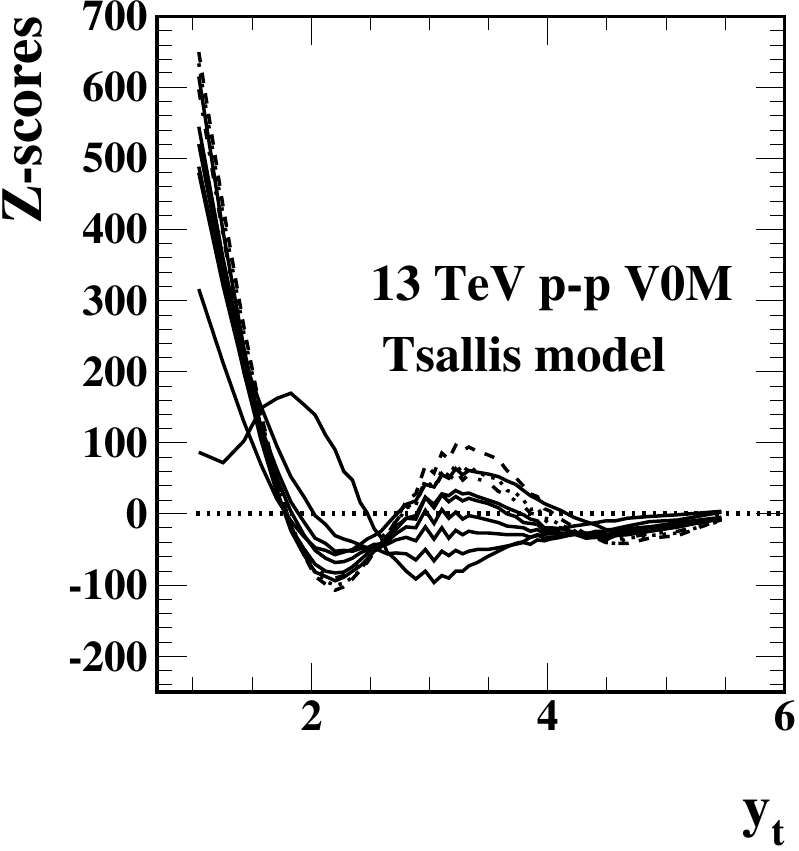}
\put(-145,99) {\bf (a)}
\put(-24,99) {\bf (b)}\\
	\includegraphics[width=1.65in,height=1.6in]{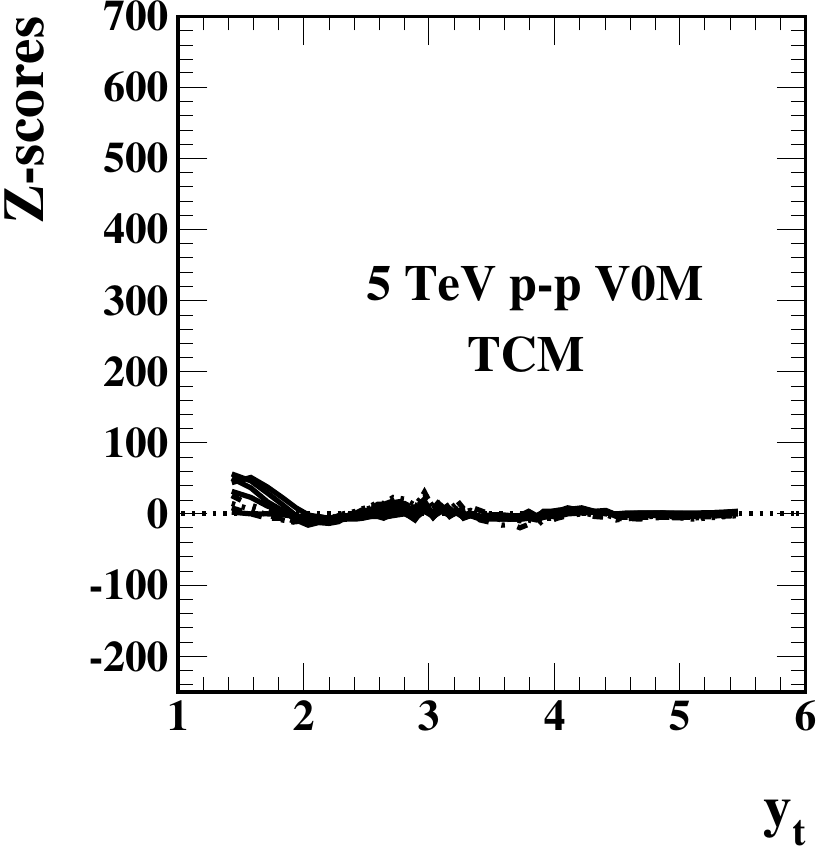}
\includegraphics[width=1.65in,height=1.6in]{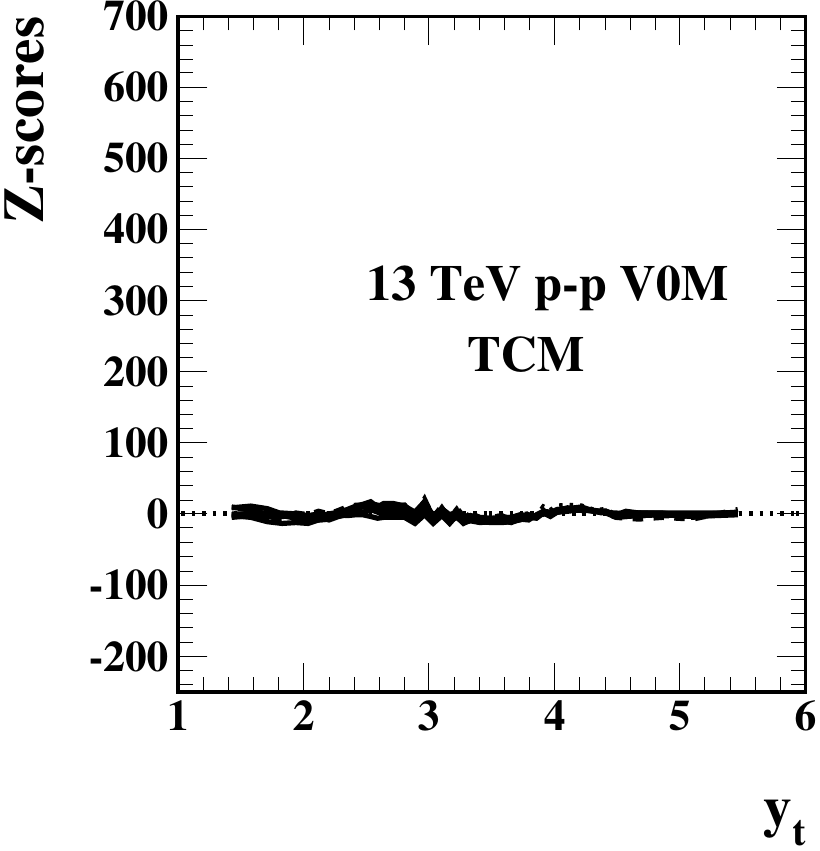}
\put(-145,99) {\bf (c)}
\put(-24,99) {\bf (d)}\\
	\caption{\label{compare}
		data-model differences in ratio to statistical errors (Z-scores) for 5 TeV (left) and 13 TeV (right) and for the Tsallis model (upper) and the TCM (lower).
	}  
\end{figure}

Figure~\ref{compare} (a,b) shows Z-scores corresponding to data/model ratios in Fig.~\ref{tsallisfits} (b,d).  In this format it becomes clear that the Tsallis model (a,b) dramatically fails ($|\text{Z-score}| \gg 1$) to describe data below \yt\ = 4 ($p_t \approx 4$ GeV/c) {\em where a large fraction of jet fragments appears} [see Figs.~\ref{tcm5} and \ref{tcm13} (b,d)]~\cite{eeprd,jetspec2,fragevo,ppbpid}. 

Figure~\ref{compare} (c,d) shows corresponding Z-scores for the TCM with varying hard component $\hat H_0(y_t;n_{ch})$ as described in Sec.~\ref{tcmfit}. Panel (d) repeats Fig.~\ref{v0mfit} (c) with omission of event class 9. Panel (c) is the corresponding result for 5 TeV also omitting class 9.  The 5 TeV V0M spectrum data were treated the same as 13 TeV data to obtain 5 TeV  $\hat H_0(y_t;n_{ch})$ parameters. The deviations from zero for 5 TeV closely follow those for 13 TeV, buttressing a conjecture that the TCM Z-score structure may indicate imperfect efficiency correction or distorted \pt\ values. What should be clear from this comparison is the dramatic difference between TCM Z-scores $O(10)$ compared to Tsallis Z-scores approaching 1000.

\subsection{Summary}

As noted above, Ref.~\cite{cleymans} seems to offer contradictory observations about the fit quality of the Tsallis model. In the Summary appears ``The non-extensive Tsallis distributions leads [sic] to a {\em very good description}  [emphasis added] of experimental data for the complete transverse momentum range in various multiplicity classes and collision energies.'' But elsewhere one finds ``In case of the lowest multiplicity class...the  Tsallis  distribution  function  seems  to  do  a  good  job  in  describing the  charged  particle  spectra.   On  the  other  hand,  for the highest multiplicity class...the Tsallis description of the spectra becomes  worsen [sic].   Although  it  is  seen  that  the  multiplicities [are those] achieved in heavy-ion collisions, Tsallis distribution function completely fails to describe the data....  We have seen the onset of collectivity in high-multiplicity pp collisions at the LHC energies. Tsallis distribution doesn't account for this, which could be the reason of the above observation.'' Figure~\ref{compare} suggests that the Tsallis model completely fails to describe \pp\ \pt\ spectrum data.

Within the Tsallis approach to spectrum modeling parameter $q$ measures the {\em statistical nonuniformity} of a many-body system: ``The deviation of the non-extensive parameter, $q$ from unity tells about the departure of the system from thermodynamic equilibrium....'' However, the same parameter may be confused with other basic issues in high-energy nuclear collisions: The increase of $q$ with collision energy ``might be understood as being due to...contributions from...hard scatterings....'' It is observed that $q-1$ increases with \nch\ and then saturates at a constant value. That trend ``...could be because of jet-fragmentation contributing to the particle production making produced high-$p_T$ particles come out of the system without interaction.'' On the other hand,  in  the limit $n_{ch} \rightarrow 0$ the Tsallis model must approximate soft component $\hat S_0(y_t)$ (the jet-related hard component becomes negligible), suggesting the proper interpretation of Fig.~\ref{qt} (left): For low \nch\ $q$ approaches a value corresponding to soft-component $\hat S_0(y_t)$ whereas for high \nch\  $q$ approaches a value corresponding to hard-component $\hat H_0(y_t)$. 

In effect, the single Tsallis model  is attempting to accommodate two {\em nearly-fixed} data components. Its parameters must then vary dramatically as in Fig.~\ref{qt} but the data description is nevertheless poor. In contrast, the TCM with two model components requires only modest {\em interpretable} parameter variation (Sec.~\ref{tcmfit}), and the spectrum data are described within their uncertainties.

Commenting on the $q$ vs \nch\ trends in Fig.~\ref{qt} (left) Ref.~\cite{cleymans} states ``The  non-extensive parameter is higher for the 13 TeV and this can be understood as the contributions from the hard scatterings (jet contribution) in pp collisions at $\sqrt{s} = 13$ TeV is [sic] larger than  at  5.02  TeV.'' ``Larger contributions'' might be understood to mean more jets, but an increase in jet {\em number} does not necessarily change the power-law ``power,'' as illustrated in Fig.~\ref{v0mfit} where the number of jets increases dramatically with \nch\ but for V0M selection the slope of the exponential tail remains constant. What does change with collision energy is the underlying minimum-bias jet spectrum as described in Ref.~\cite{jetspec2}, which then controls the hard-component tail structure as described in Ref.~\cite{fragevo}.

\section{Blast-wave spectrum model} \label{bwmodel}

Reference~\cite{cleymans} also applies a blast-wave model to \pt\ spectrum data for 5 and 13 TeV V0M \pp\ spectra.
Compared to the Tsallis model the BW model is said to be a ``more standard description'' and is ``based on collective flow in small systems.'' The BW model is said to be ``quite good in explaining the bulk part [below 2.5 GeV/c] of the system, however it fails at low-$p_T$ {below 0.5 GeV/c} which could possibly be due to the decays of hadronic resonances.'' It is asserted that ``The applicability of BGBW [BW] model is verified by fitting the transverse momentum spectra of the bulk part ($\sim$ 2.5 GeV/c) for both 5.02 and 13 TeV energies and also in different multiplicity classes.''  The paper summary contains reference to ``collectivity seen in these events.''

\subsection{Blast-wave method}

The BW model invoked to fit spectrum data in Ref.~\cite{cleymans} is nominally adopted from Ref.~\cite{ssflow} that introduced a BW model to describe pion spectra from 200 GeV fixed-target \mbox{S-S} collisions at the SPS. The relevant formula is Eq.~(7) (second line) of Ref.~\cite{ssflow}
\bea \label{bweq}
\frac{dn}{m_t dm_t} \hspace{-.05in} &\propto & \hspace{-.05in} m_t \hspace{-.05in} \int_0^R \hspace{-.1in} r dr I_0\left[\frac{p_t \sinh(\rho)}{T}\right] K_1\left[\frac{m_t \cosh(\rho)}{T}\right],~~~~
\eea
with $\rho = \tanh^{-1}(\beta)$ and $\beta(r) = \beta_s(r/R)^m$, where $m = 1$ corresponds to Hubble expansion, $\beta_s$ is the surface expansion speed and $I_0$ and $K_1$ are modified Bessel functions. In effect, Eq.~(\ref{bweq}) represents a thermal (exponential) \mt\ spectrum in the boost frame convoluted with a boost distribution on particle source radius to describe the spectrum observed in the lab frame~\cite{quadspec}.  The corresponding BW spectrum model given in Eq.~(14) of Ref.~\cite{cleymans} is
\bea \label{badbw}
\frac{d^2N}{dp_t dy} &=& D \int_0^{R_0} \hspace{-.1in} r dr K_1\left[\frac{m_t \cosh(\rho)}{T}\right] I_0\left[\frac{p_t \sinh(\rho)}{T}\right]~~~
\eea
that is missing factors $p_t m_t$ compared to Eq.~(\ref{bweq}).

In the present analysis the form given by Eq.~(\ref{badbw}) with parameter values for $\langle \beta_t \rangle$ and $T_{kin}$ as given in Fig.~\ref{dumb} below (Figs.~11 and 12 of Ref.~\cite{cleymans}) was initially applied to the \pp\ spectrum data for V0M event selection as shown in Sec.~\ref{tcmspecdat}. The model was observed to deviate strongly from data. Two changes were required: (a) restore the missing factors $p_t m_t$ appearing in Eq.~(\ref{bweq}) and (b) increase the values for $T_{kin}$ appearing in Fig.~12 of Ref.~\cite{cleymans} by factor 2.15. See Fig.~\ref{dumb} and associated text below. With those changes the results in Fig.~\ref{bwfits} (a) below are comparable to those in Fig.~9 of Ref.~\cite{cleymans}.

\subsection{Blast-wave results}

\begin{figure}[h]
	\includegraphics[width=3.3in,height=1.6in]{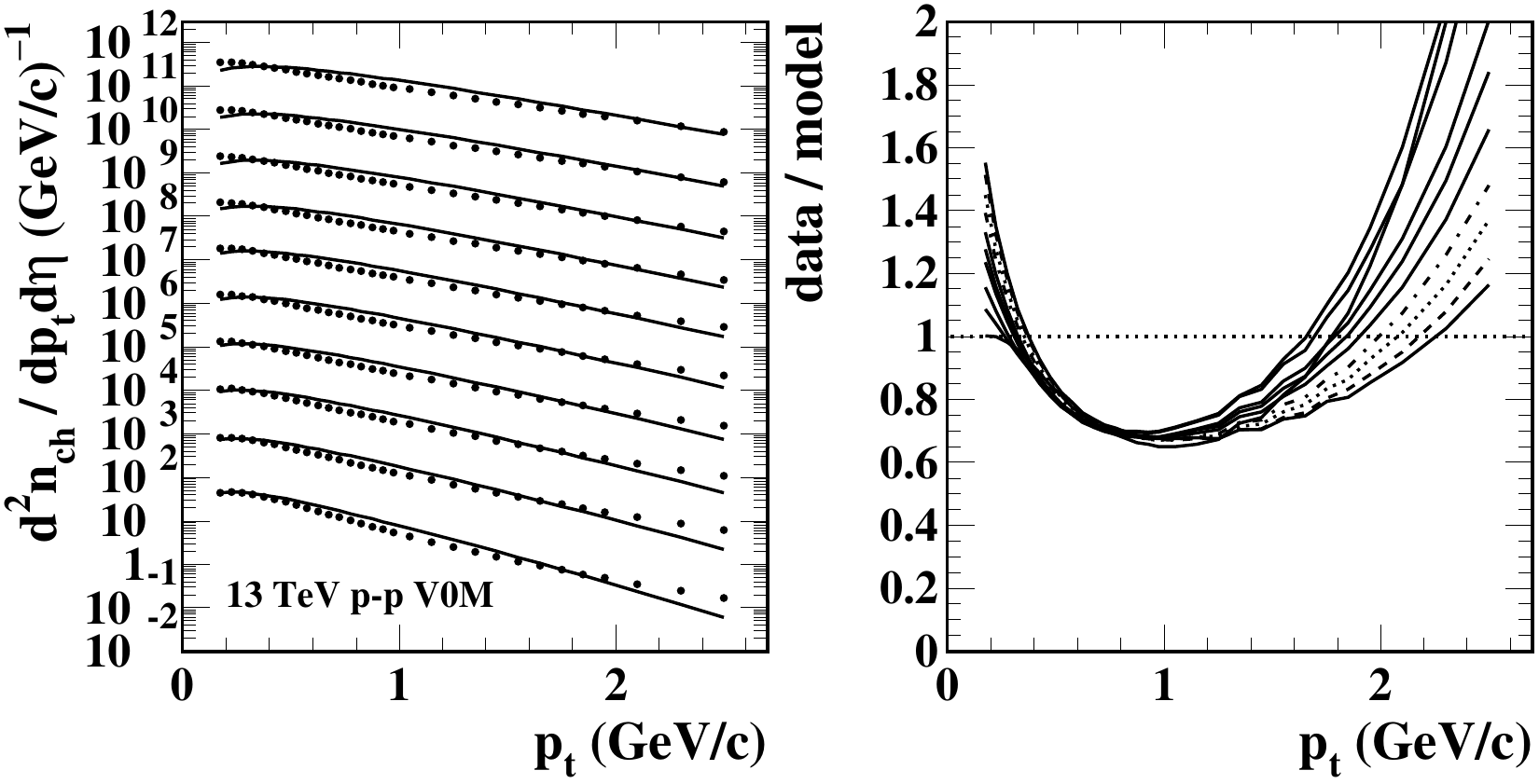}
\put(-137,100) {\bf (a)}
\put(-34,100) {\bf (b)}\\
	\includegraphics[width=1.65in,height=1.6in]{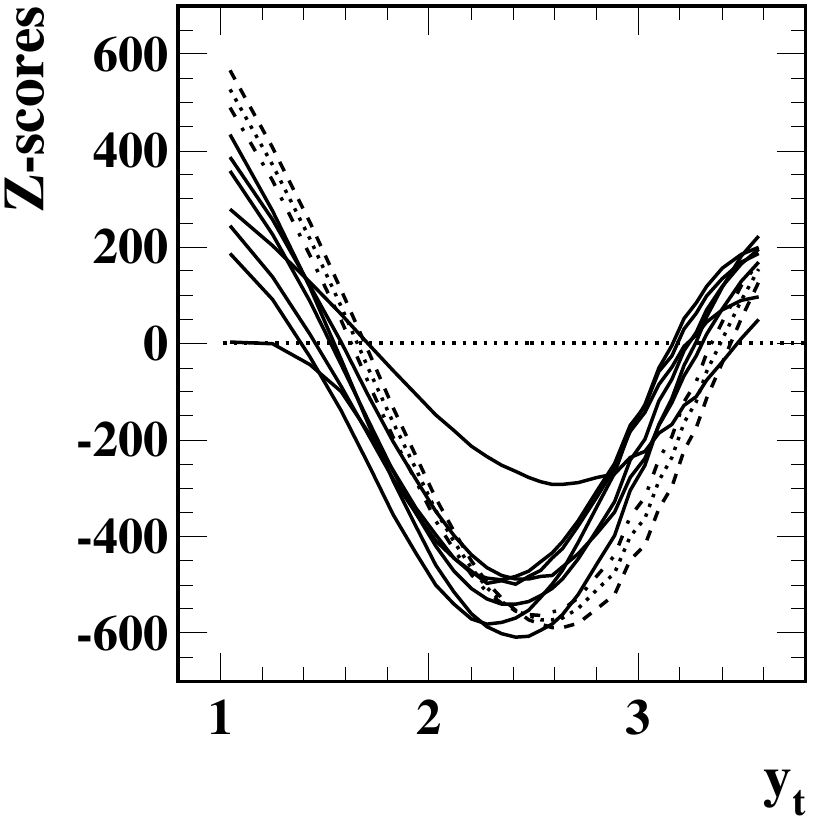}
	\includegraphics[width=1.65in,height=1.6in]{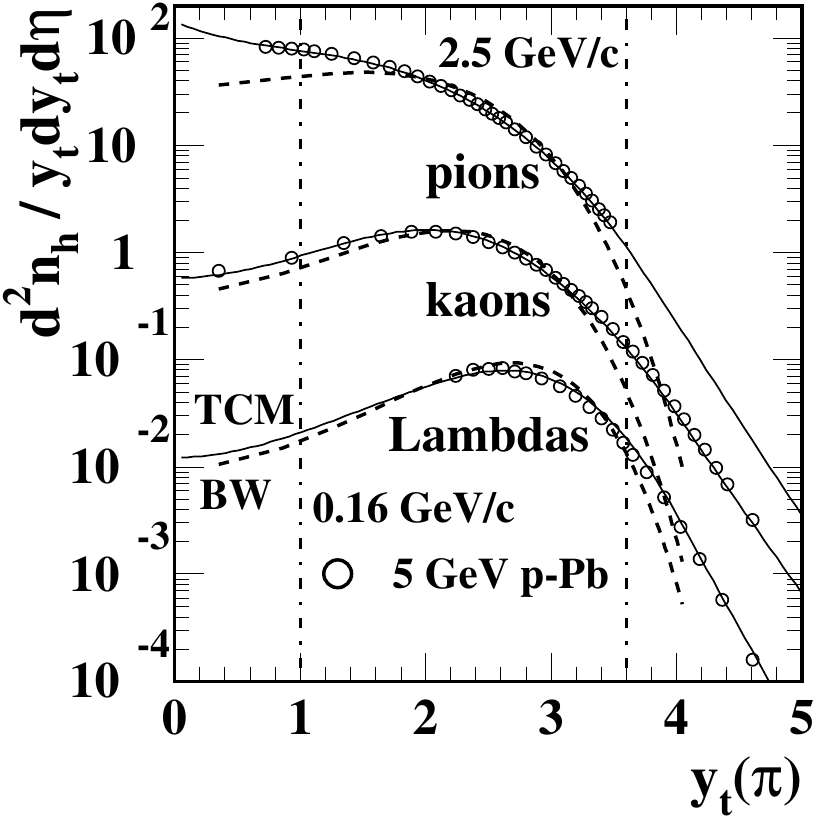}
\put(-145,99) {\bf (c)}
\put(-24,99) {\bf (d)}
	\caption{\label{bwfits}
		(a) Blast-wave model fits (curves) and spectrum data (points) for V0M event selection from 13 TeV \pp\ collisions.	
		(b) Data/model ratios for panel (a).
		(c) Z-scores for panel (a).
		(d) Individual BW models for pions, kaons and protons (dashed) in relation to PID spectra from peripheral \ppb\ events (class 6 with $\bar \rho_0 \approx 9.7$) with mean charge density approximating class VII \pp\ events with $\bar \rho_0 \approx 8.46$. Also shown are TCM results for that system (solid) from Ref.~\cite{ppbpid}. Spectra are multiplied by 1, 5 and 20 to provide separation.
		Dash-dotted lines indicate \yt\ values for 0.16 and 2.5 GeV/c.
	}  
\end{figure}

Figure~\ref{bwfits} (a) shows BW model curves with those conditions (solid) in relation to 13 TeV \pp\ V0M spectrum data (points) from Ref.~\cite{alicenewspec}. The results appear similar to what is presented in Fig.~9 of Ref.~\cite{cleymans}, but even within the restricted \pt\ fit interval there are substantial data-model deviations. The data-model relation is essentially the same for 5 TeV data.  Figure~\ref{bwfits} (b) shows data/model ratios with large deviations from 1 comparable to or exceeding those for the Tsallis model in Fig.~\ref{tsallisfits}.  Figure~\ref{bwfits} (c) shows Z-scores comparable to or exceeding those from the Tsallis model in Fig.~\ref{compare} (a,b). It should be of interest to understand how those deviations arise.

Figure~\ref{bwfits} (d) shows identified-hadron (PID) data (points) from 5 TeV \ppb\ collisions~\cite{aliceppbpid} with a multiplicity low enough (class 6, $\bar \rho_0 \approx 9.7$) that those peripheral \ppb\ collisions are effectively equivalent to single \pn\ collisions~\cite{ppbpid,tomglauber}. The solid curves are the corresponding TCM from Ref.~\cite{ppbpid} that describes all \ppb\ data accurately. It is especially notable that the TCM describes $K^0_\text{S}$ (kaons) data from 7 GeV/c {\em down to zero \pt}. The dashed curves are BW model curves for individual hadron species that, when summed with weights $w_i$, correspond to the V0M class VII ($\bar \rho_0 \approx 8.46$) curve in Fig.~\ref{bwfits} (a). The BW curves are rescaled to correspond with $\bar \rho_0 \approx 9.7$.

\begin{figure}[h]
	\includegraphics[width=3.3in,height=1.6in]{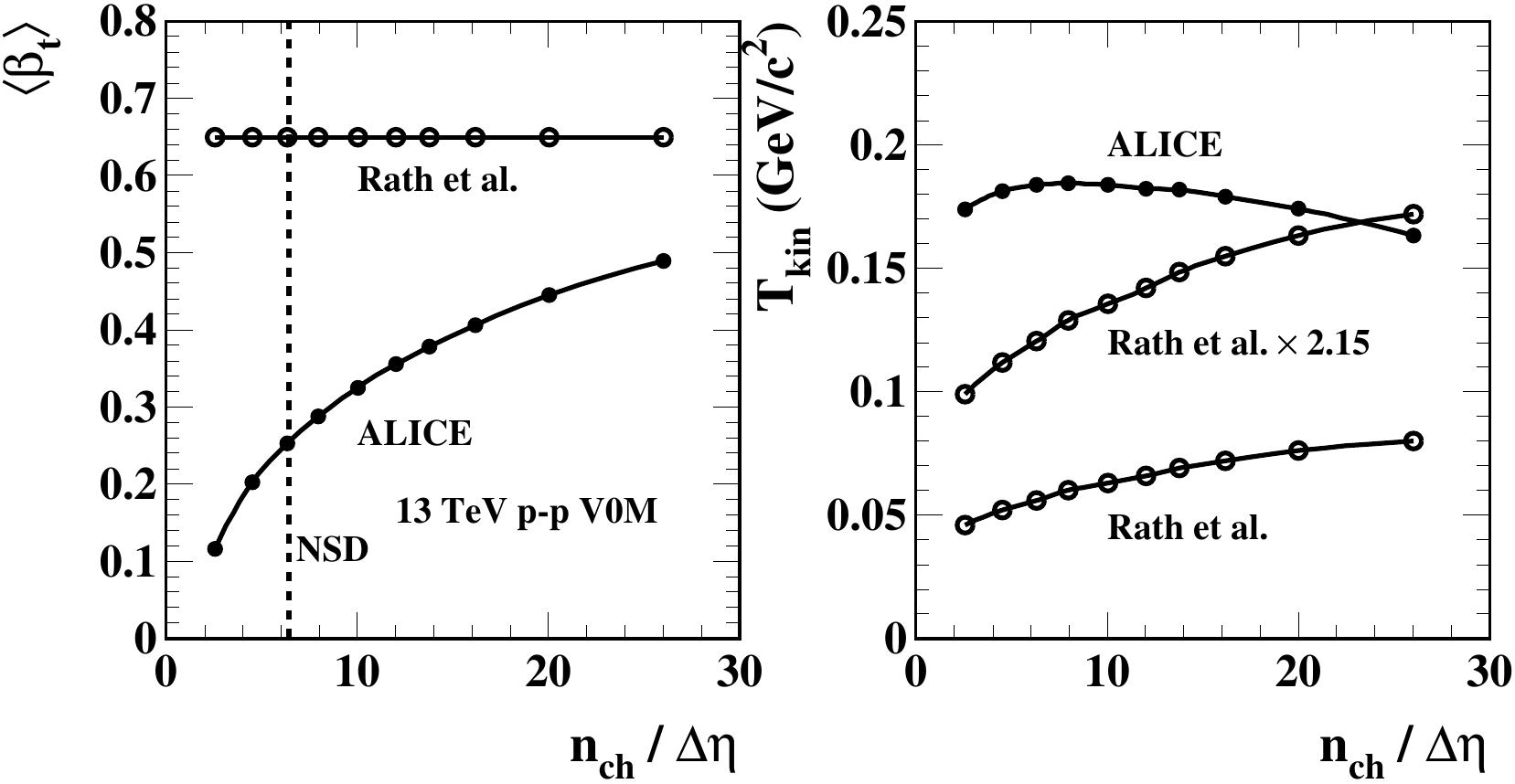}
	\caption{\label{dumb}
		Left: Mean transverse speed $\langle \beta_t \rangle$ vs charge density $\bar \rho_0$ for Ref.~\cite{cleymans} (hatched band) and Ref.~\cite{aliceppbpid} (solid points and curve).
		Right: Kinetic freezeout parameter $T_{kin}$ vs  charge density $\bar \rho_0$  for Ref.~\cite{cleymans} (lower open points and curve) and Ref.~\cite{aliceppbpid} (solid points and curve). The upper open points and curve are explained in the text.
	}  
\end{figure}

Figure~\ref{dumb} shows parameters $\langle \beta_t \rangle$ and $T_{kin}$ for 13 TeV \pp\ collisions with V0M event selection from Ref.~\cite{cleymans} Figs.~11 and 12 respectively (Rath {\em et al.}). Also shown are results for the same collision system from Ref.~\cite{alicepppid} (ALICE). The ALICE results for \pp\ are approximately consistent with those for \ppb\ and \pbpb\ in terms of \nch\ dependence. The parameter trends from Ref.~\cite{cleymans} are thus quite exceptional compared to other analyses and collision systems. However, that does not mean any results from BW analysis reflect manifestations of a flow phenomenon. The same data from Ref.~\cite{alicepppid} are described in Sec.~\ref{tcmfit} by the TCM within data uncertainties, and the TCM includes only longitudinal proton dissociation and transverse jet formation as physical mechanisms.

\subsection{Summary}

The large BW data-model deviations can be explained as follows: As noted, the BW model in Eq.~(\ref{bweq}) is equivalent to convoluting a thermal exponential in the boost frame (the particle source frame) with a boost distribution modeling (in this case) a Hubble-expanding (flowing) source distribution to obtain the BW model in the lab frame where particles are detected. The boost implies that the model {\em must be suppressed} at lower \pt\ relative to an unboosted source and enhanced at higher \pt\ because of the boost (a boost is simply a translation on \yt). The boost effect is most apparent (on \pt) for Lambdas with large mass. The BW model must also fall increasingly below data at higher \pt\ because there is no jet component. Cutting off the  fit interval merely by assuming that data below some \pt\ value correspond exclusively to a locally thermalized source is not justified. The majority of jet fragments appears there, as observed in spectrum data (see Sec.~\ref{tcmspecdat}) and predicted by pQCD (see Ref.~\cite{fragevo}).

Thus, an inevitable result of fitting the BW model to \pp\ spectra is the Z-score trends of Fig.~\ref{bwfits} (c): the model falls well below the data at lower \pt\ and higher \pt\ for well-understood reasons but exceeds the data at intermediate \pt\ to minimize $\chi^2$. Attempts to fit the BW model to spectrum data confront rapid increase (quadratic dependence on \nch) of jet production, with fragment-density maximum near 1 GeV/c. Of two model parameters $\langle \beta_t \rangle$ should increase rapidly to accommodate the increasing amplitude of the hard component (with its fixed mode), and $T_{kin}$ may or may not vary to accommodate data-model differences at lower \pt. The results from Ref.~\cite{cleymans} are in dramatic conflict with other recent analysis (e.g.\ Ref.~\cite{qgpreview}, Fig.~3, and Ref.~\cite{alicepppid}, Fig.~4). Whereas the BW $T_{kin}$ model parameter for 13 TeV \pp\ collisions from Ref.~\cite{cleymans} varies over the interval 45 -- 80 MeV, TCM soft-component slope parameter $T$ remains fixed near 145 MeV for all collision systems from 17 GeV to 13 TeV.


Reference~\cite{cleymans} presents the following conclusions: ``The collective radial flow velocity, $\langle \beta_t \rangle$ is almost independent of the collision energy and multiplicity classes. The kinetic freeze-out temperature, $T_{kin}$ however, shows a clear dependence on the multiplicity classes.'' In contrast, Ref.~\cite{qgpreview}  (a recent review of QGP-related LHC results) observes in regard to its Fig.~3 (presenting ALICE BW fit parameters) ``For small systems pp and p-Pb, the $T_{kin}$ remains constant while $\langle \beta_T \rangle$ increases rapidly with multiplicity. At similar multiplicity values, the $\langle \beta_T \rangle$ is larger for small systems.''  Reference~\cite{cleymans} summarizes: ``Conclusively, the BGBW [BW model] explains the bulk part of the transverse momentum spectra and the description is better for the higher multiplicity classes....'' But  the ``bulk part'' is where the great majority of jet fragments appears as the spectrum hard component (see Sec.~\ref{tcmspecdat}).

\section{Systematic uncertainties} \label{syserr}

It is customary to include a separate article section that evaluates the uncertainty of the various methods, results and conclusions derived from a statistical study of data. However, systematic uncertainties have been a central topic throughout the present article. Two major issues do emerge: the distinction between data/model ratios and Z-scores as established by Eq.~(\ref{suppress}) and the persistent presence of systematic distortions in spectrum data.

\subsection{Spectrum ratios vs $\bf \chi^2$ tests}

\begin{figure}[h]
	\includegraphics[width=1.65in,height=1.63in]{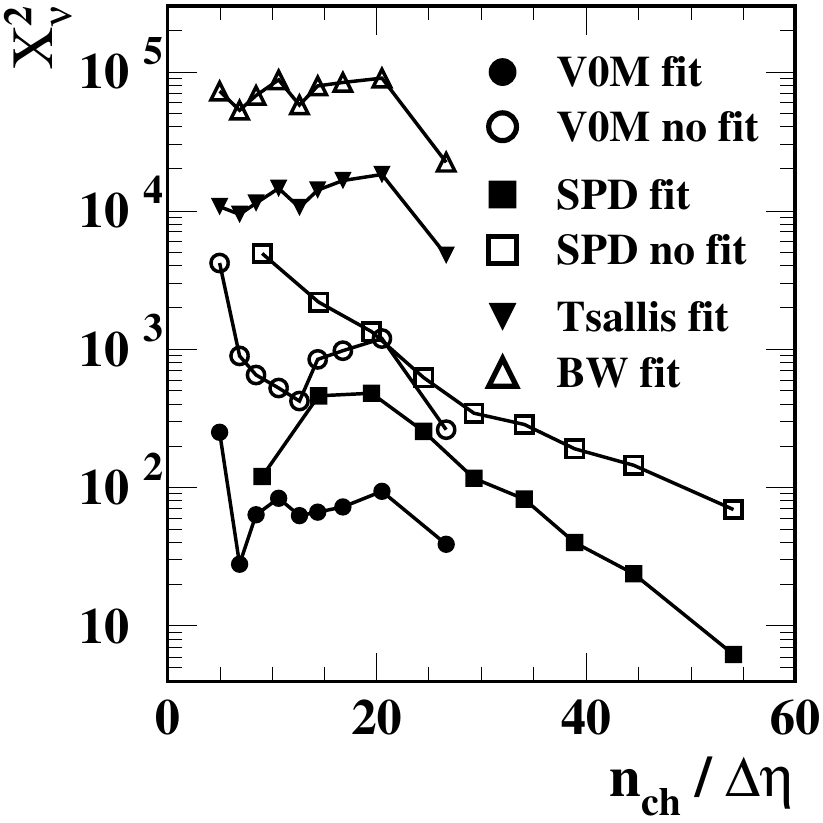}
	\includegraphics[width=1.65in,height=1.65in]{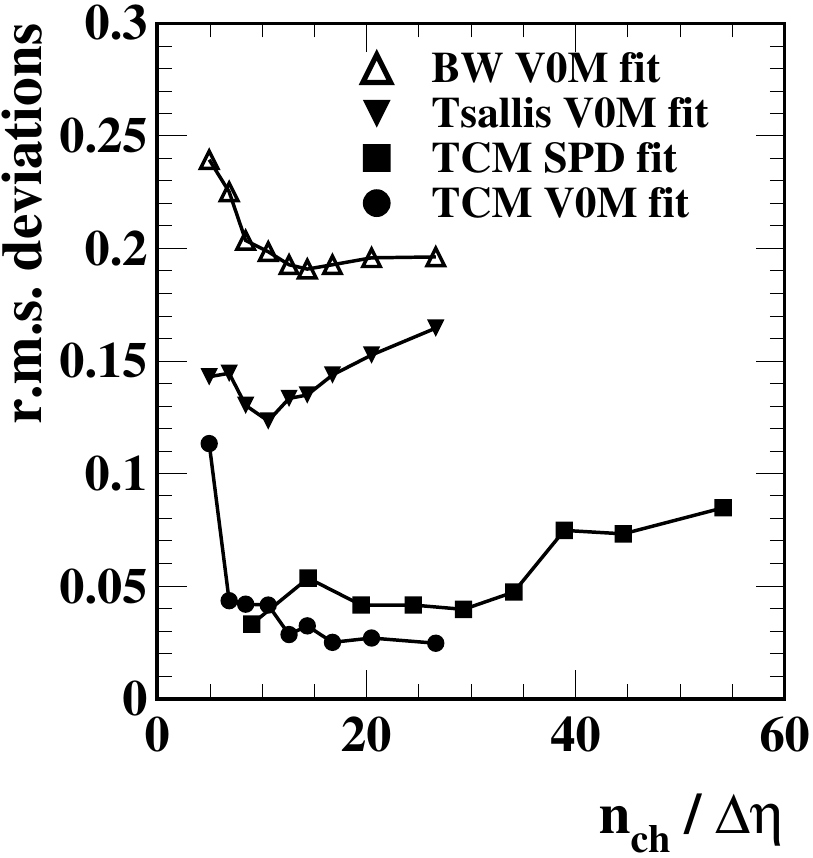}
	\caption{\label{rms}
	Left: $\chi^2_\nu$ defined by Eq.~(\ref{chinu}) for the TCM applied to V0M and SPD spectra with parameters adjusted for \nch\ classes (fit) and held fixed (no fit), and for Tsallis and BW fits.
	Right: Deviations (r.m.s.) from 1 denoted by the left-hand side of Eq.~(\ref{suppress}) for the TCM applied to V0M and SPD spectra and for Tsallis and BW models applied to V0M spectra.
	} 
\end{figure}

Figure~\ref{rms} (left) shows $\chi^2_\nu$ values defined by Eq.~(\ref{chinu}) evaluating the fit quality of three models applied to 13 TeV \pp\ \pt\ spectra with V0M and SPD event selection. The TCM is applied both as a fixed reference model (open circles and squares) and as a model with parameters varied to accommodate different \nch\ classes and selection methods (solid dots and squares) that can be considered a ``fit'' of sorts. Results from Tsallis (solid triangles) and BW (open triangles) models for V0M data are also shown.

The general trends can be interpreted as follows: (a) Results for V0M event selection are approximately constant in magnitude vs \nch\ with a few exceptions. Data/model ratio deviations from 1 tend to be similar across \nch\ classes, and the error/data ratios for V0M in Fig.~\ref{errorrats} (left) are also similar. (b) The TCM $\chi^2_\nu$ trends for SPD event selection decrease strongly with \nch\ due to the very strong variation of error/data ratios in Fig.~\ref{errorrats} (right). (c) Given the large difference in error/data trends for V0M and SPD the results for ``fitted'' TCM are comparable for the two selection methods. (d) Even the fixed-reference (no fit) TCM is {\em one to two orders of magnitude} lower than the $\chi^2_\nu$ values for Tsallis and BW models, suggesting falsification of those models.

Figure~\ref{rms} (right) shows r.m.s.\ averages for $ \text{data/model} -1$ on the left side of Eq.~(\ref{suppress}).  The largest difference between data trends is a factor 10 between TCM fits for V0M (solid dots) vs BW fits for V0M (open triangles). That might be expected to correspond to a factor 100 difference for the squared quantity $\chi^2_\nu$ in the left panel, but the actual factor is approximately 1000 ($\approx 80$ vs $\approx 80,000$). The difference arises because in  Eq.~(\ref{suppress}) $ \text{data/model} -1$ on the left is equal to the product of Z-scores (the source of $\chi^2_\nu$ via Eq.~\ref{chinu}) and the error/data ratio. The r.m.s.\ values of the product are then suppressed by factors 10-30 compared to what might be expected from $\chi^2_\nu$ values on the left.  In plots of spectrum ratios, for instance Fig.~9 (lower) of Ref.~\cite{cleymans}, an r.m.s.\ average deviation of 0.2 seems small. But it corresponds to very large $\chi^2_\nu$ values as in Figure~\ref{rms} (left) that, according to standard statistical practice, should overwhelmingly falsify the BW model.

There is a qualitative difference between TCM data-model differences and those for Tsallis and BW models. In Figs.~\ref{tsallisfits} (b,d) and \ref{bwfits} (b) the data/model ratios exhibit long-wavelength, large-amplitude deviations that suggest fundamental differences between model and data. In Figs.~\ref{v0mfit} (b) and \ref{spdfit} (b) the TCM-data deviations are short-wavelength and very similar for most \nch\ classes and both selection methods. As suggested above, it is possible that the small-scale or local structures reflect systematic distortions injected into final data during processing of tracking data and that $\chi^2_\nu$ and r.m.s.\ values for the TCM fitted model in Fig.~\ref{rms} are only upper bounds.

\subsection{Systematic distortions} \label{sysdistort}

In the analyses of Secs.~\ref{tcmfit}, \ref{tsallismodel} and \ref{bwmodel} the errors $\sigma_i$ assumed are those shown in ratio to data by curves in Fig.~\ref{errorrats}. However, the Z-scores for the TCM applied to V0M and SPD data in Figs.~\ref{v0mfit} (c) and \ref{spdfit} (c) shows the same pattern of short-wavelength systematic distortions for V0M and SPD and for every \nch\ class, suggesting that the same fractional systematic distortion was applied to every \pp\ \pt\ spectrum in the process of generating particle data from tracking data. In that case uncertainty $\sigma_i$ in Eq.~(\ref{zscore}) may be substantially underestimated, at least at lower \yt. Based on TCM V0M r.m.s.\ trends in Fig.~\ref{rms} (right) a fixed value $\sigma_{i,sys} = 0.02$ (dashed lines in Fig.~\ref{errorrats}) is added in quadrature to $\sigma_{i,stat}$ (curves in Fig.~\ref{errorrats}), and Z-scores are recomputed with those $\sigma_i$ values.

\begin{figure}[h]
	\includegraphics[width=1.65in,height=1.6in]{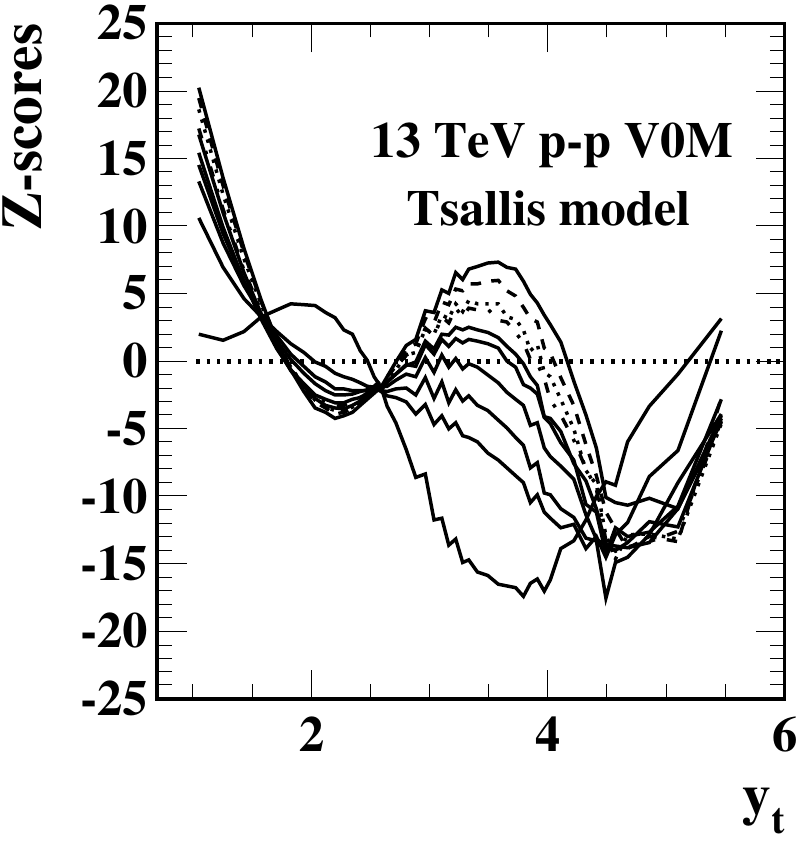}
	\includegraphics[width=1.65in,height=1.6in]{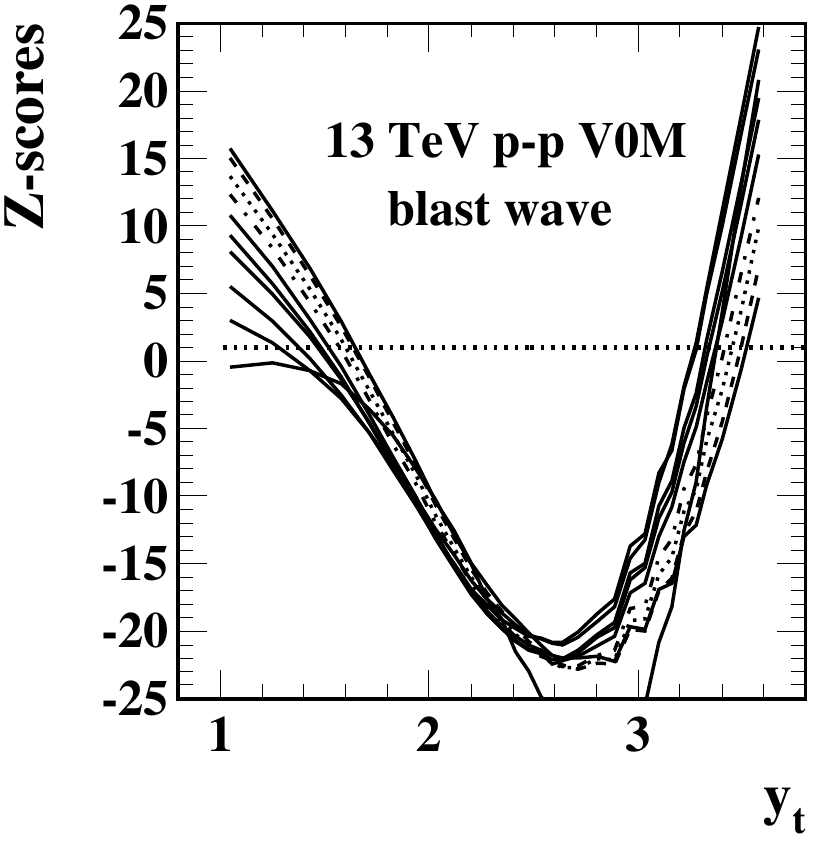}
	\put(-145,102) {\bf (a)}
	\put(-25,102) {\bf (b)}\\
	\includegraphics[width=1.65in,height=1.6in]{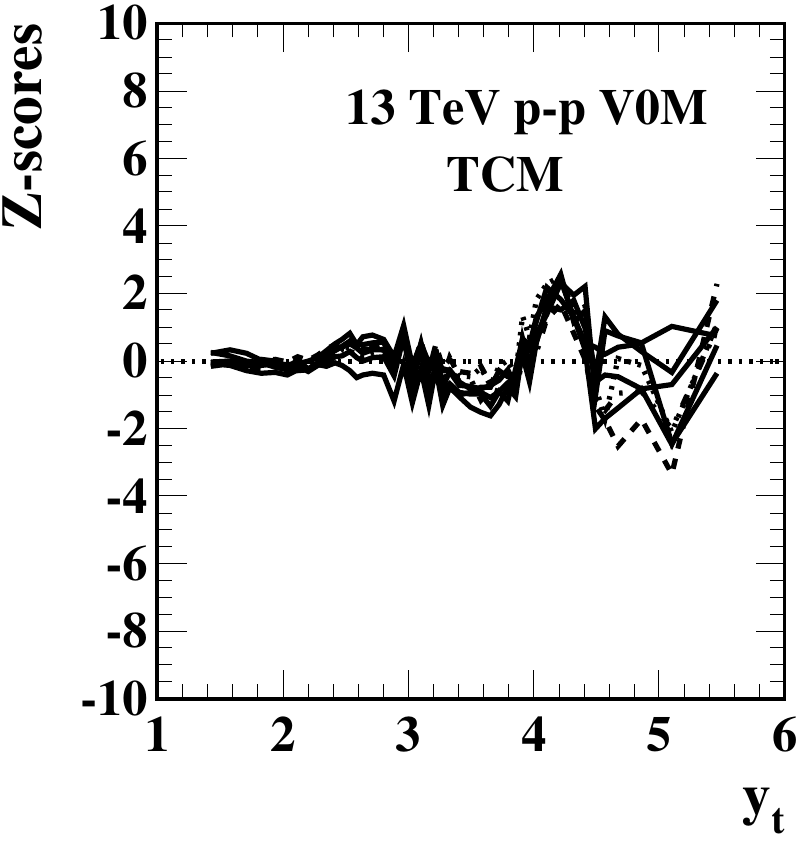}
	\includegraphics[width=1.65in,height=1.6in]{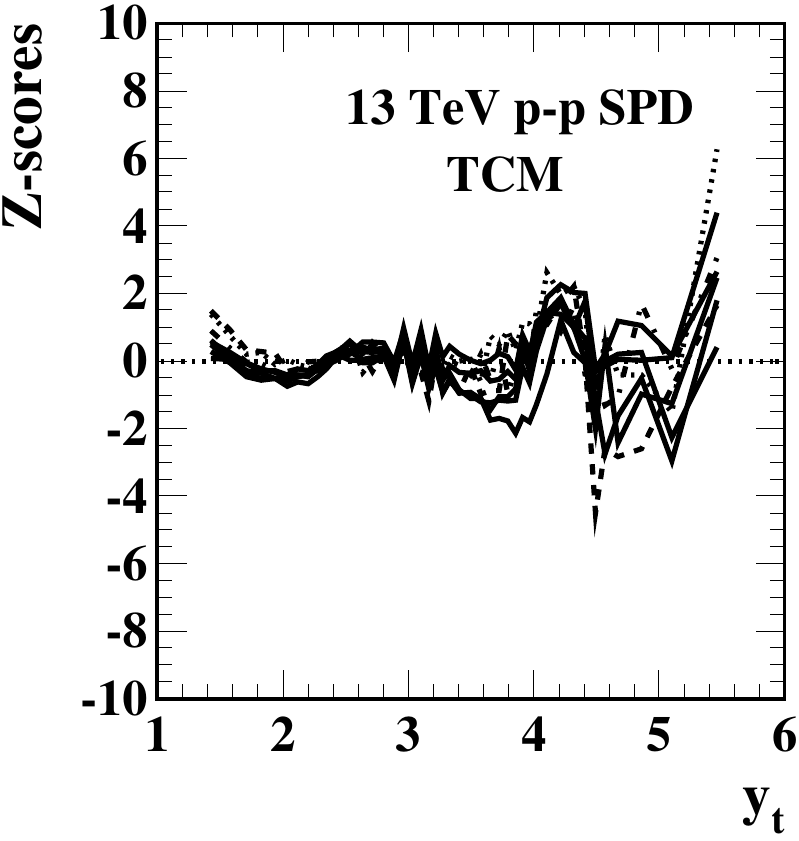}
	\put(-145,90) {\bf (c)}
	\put(-27,90) {\bf (d)}\\
	\caption{\label{systematic}
Z-scores for 13 TeV \pp\ collisions assuming a quadratic combination of statistical $\sigma_{i,stat}$ and systematic $\sigma_{i,sys}$ errors in Eq.~\ref{zscore} and for (a) Tsallis model and (b) blast-wave model applied to V0M data and for (c) TCM applied to V0M data and (d) TCM applied to SPD data.
	} 
\end{figure}

Figure~\ref{systematic} (a) and (b) show recomputed Z-scores for Tsallis and BW models respectively. Compared to the results in Figs.~\ref{compare} (a,b) and \ref{bwfits} (c), derived using published statistical errors only, the Z-scores at lower \yt\ are reduced by more than a factor ten as expected.

Figure~\ref{systematic} (c) and (d) show corresponding results for the TCM applied to V0M and SPD data. To improve visibility of relevant structure event classes $n=9$ from (c) and $n=1$ from (d) have been omitted. The former is motivated by the inability of the simple TCM model to deal with the strong bias for that low V0M \nch\ class with $\bar \rho_0 = 4.94$ which falls below the NSD value 6.4. The latter is motivated by the large statistical errors of the highest SPD \nch\ class with $\bar \rho_0 \approx 54$ due to small event number, as indicated by the uppermost curve in Fig.~\ref{errorrats} (right). It is clearly apparent that the same distortion pattern appears in every spectrum. Statistical fluctuations superposed on that pattern are far smaller than the pattern mean values themselves below \yt\ = 4.5, in accord with Fig.~\ref{errorrats}. Given the consistent distortion patterns in Fig.~\ref{systematic} (c) and (d) it is possible to correct  spectrum data as described below.
In any event, comparing upper and lower panels of Fig.~\ref{systematic} it should be clear, even with introduction of a systematic-uncertainty element, that Tsallis and BW models are dramatically rejected by spectrum data whereas the TCM describes spectrum data within their uncertainties.

\begin{figure}[h]
	\includegraphics[width=3.3in,height=1.63in]{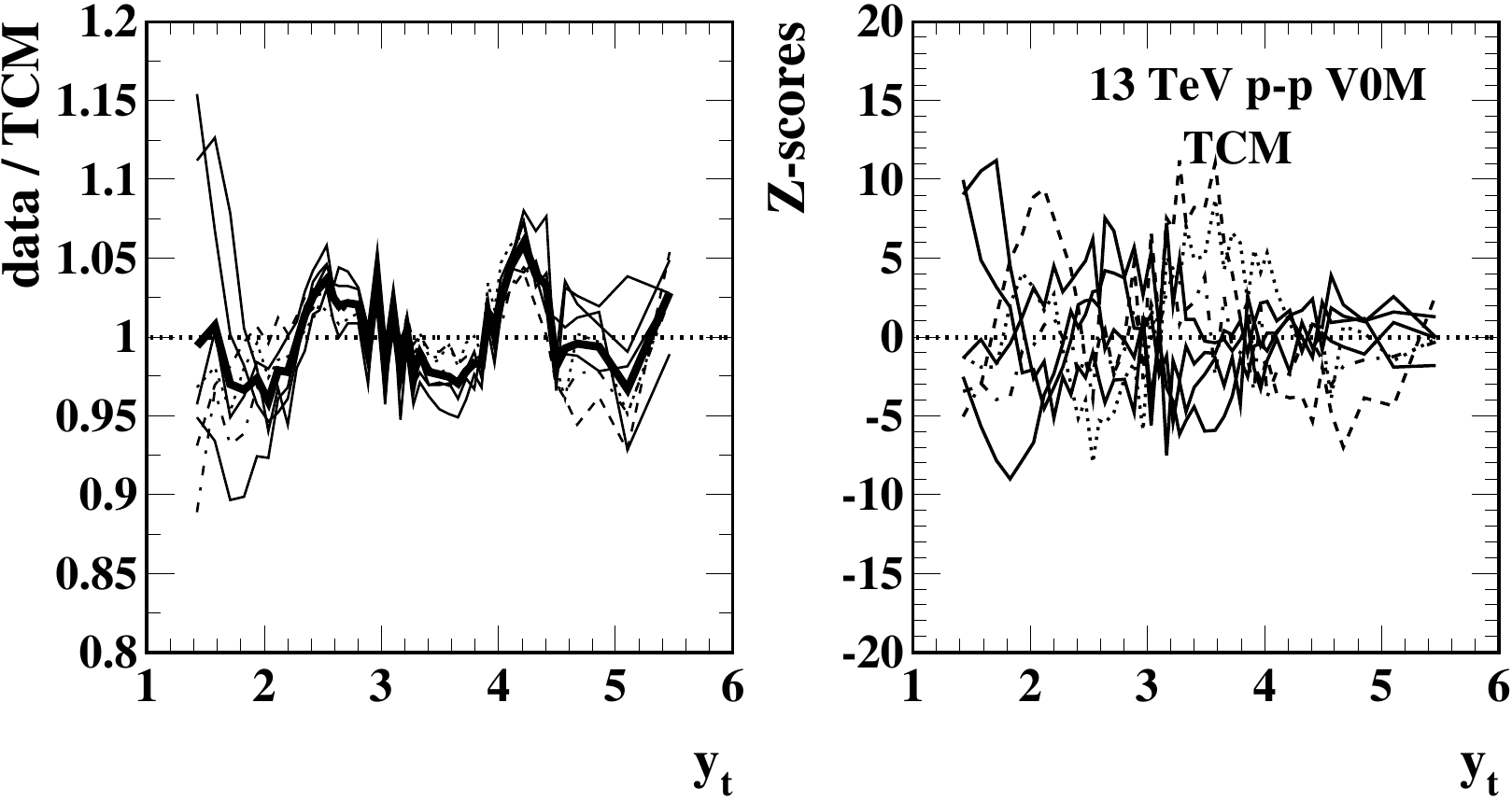}
	\caption{\label{corrected}
		Left: Data/TCM ratios for 13 TeV V0M data and for \nch\ classes $n \in [2,8]$ (curves of various line styles) and the arithmetic average (bold solid).
		Right: Z-scores calculated with statistical errors only and with data hard components divided by the bold solid curve at left (corrected). This panel can be compared with Fig.~\ref{v0mfit} (c) showing uncorrected data.
		} 
\end{figure}

Figure~\ref{corrected} (left) shows data/TCM ratios as in Fig.~\ref{v0mfit} (b) for event classes $n \in [2,8]$. The excluded classes either have increased statistical noise ($n = 1$) or include substantial physically-significant deviations from the TCM ($n = 9$). The included seven classes are arithmetically averaged (bold solid curve). All data hard components are then divided by the same bold curve as a correction.

Figure~\ref{corrected} (right) shows Z-scores corresponding to  corrected data. These Z-scores are calculated with statistical errors only. This panel may be compared with Fig.~\ref{v0mfit} (c) showing Z-scores for uncorrected data. Whereas the uncorrected data show Z-scores with systematic amplitudes $O(10)$, Z-scores for corrected data above show no systematic patterns and an r.m.s.\ value $O(4)$. The low-amplitude fine-scale structure is actually numerically consistent with expected Poisson fluctuations. It is then fair to compare that result with Z-scores in the hundreds from Tsallis and BW models as in Figs.~\ref{compare} and \ref{bwfits}. The corrected Z-scores in the right panel would be indistinguishable from a straight line at zero in Fig.~\ref{compare} (d).

\begin{figure}[h]
	\includegraphics[width=1.65in,height=1.63in]{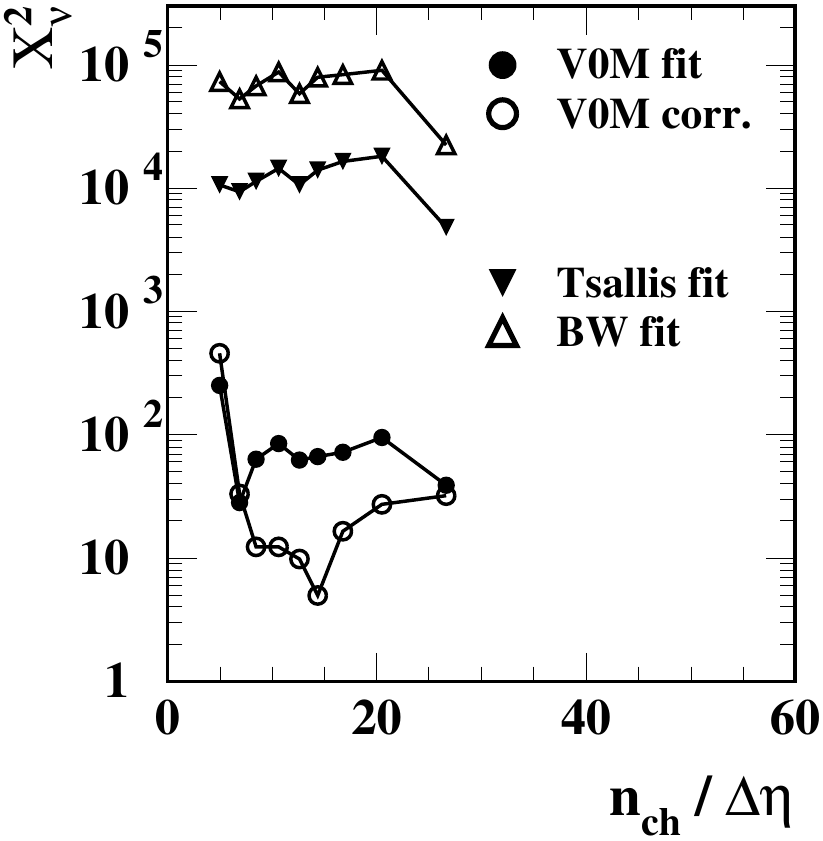}
	\includegraphics[width=1.65in,height=1.65in]{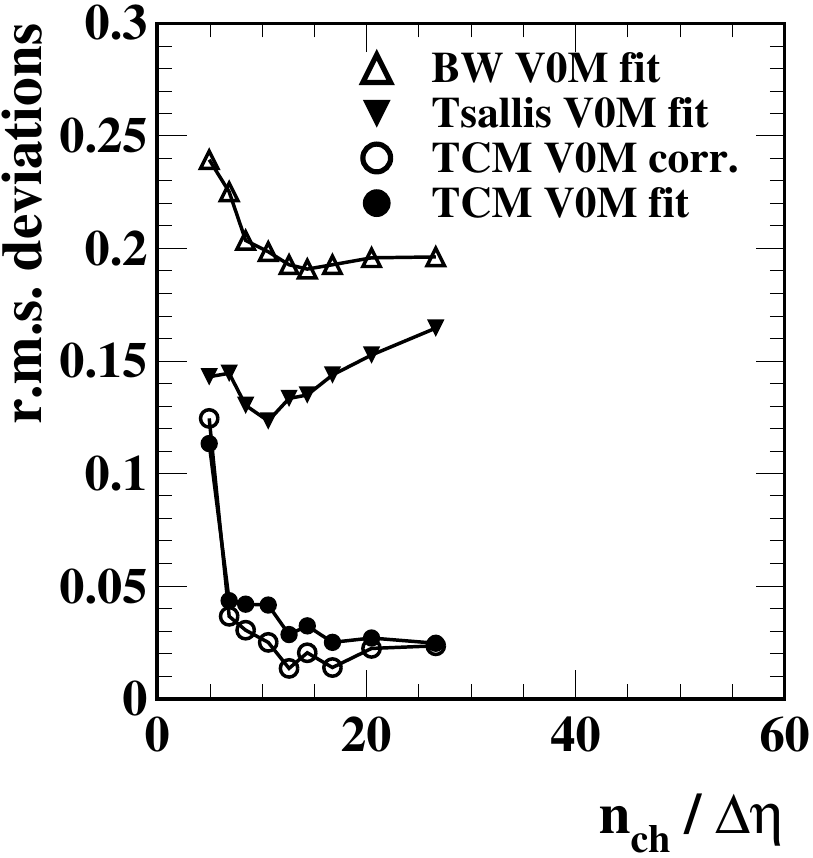}
	\caption{\label{systematic2}
	Left: $\chi^2_\nu$ defined by Eq.~(\ref{chinu}) for the TCM applied to V0M spectra with parameters adjusted for \nch\ classes (fit) and with data hard components corrected as in Fig.~\ref{systematic} (corr.), and for Tsallis and BW fits.
Right: Deviations (r.m.s.) from 1 denoted by the left-hand side of Eq.~(\ref{suppress}) for the TCM applied to V0M spectra (without and with correction) and for Tsallis and BW models applied to V0M spectra.
	}  
\end{figure}

Figure~\ref{systematic2} shows updates of Fig.~\ref{rms} illustrating the effect of corrections applied to V0M data as described above. The result for TCM fits to V0M data are roughly a factor 10 reduction in $\chi_\nu^2$ (left) and a factor 3 reduction in r.m.s.\ deviations (right). The lowest $\chi^2_\nu$ trend is consistent with an r.m.s.\ Z-score of 3-4 which is also consistent with Fig.~\ref{corrected} (right).

The residual fluctuations in Fig.~\ref{corrected} (right) may be the absolute minimum that can be expected from a real physical system. The errors $\sigma_i$ for Z-scores are predicated on statistical fluctuations following a Poisson distribution. However, real systems almost always exhibit excess or ``nonstatistical'' fluctuations that can be represented by a modified random-variable distribution --  a $q$-Gaussian represented by Eq.~(\ref{tsalliseq}) with $x/T \rightarrow  x^2/2 \sigma^2$ and with the limiting case $\exp[-x^2 / 2 \sigma^2]$ if $1/n \rightarrow 0$. For nonzero $1/n$ (a heterogeneous system) the peaked distribution exhibits raised tails away from the mode that may represent the fluctuations in  Fig.~\ref{corrected} (right). Such nonGaussian fluctuations become increasingly important with larger data volumes and relatively smaller $\sigma _i$s.

\section{Discussion} \label{disc}

In general, the fixed TCM serves as a stable and accurate reference that does not derive from fits to individual spectra. It is required to describe diverse data formats from a broad array of collision systems self-consistently. The TCM accurately separates jet and nonjet data components, greatly facilitating and simplifying data interpretation. Data-TCM deviations may reveal systematic data biases as in the present study or identify new physics beyond conventional models as in Ref.~\cite{ppquad}. The variable TCM describes \pp\ spectra within their uncertainties. In this section three topics are considered further:  (a) data information and model testing, (b) interpreting model-parameter trends and (c) collectivity in \pp\ collisions.

\subsection{Data information and model testing}

Information requires the concept of entropy as a context. Entropy can be interpreted as the logarithm of a volume (the term ``volume'' as used in a general sense to represent length, area, 3D volume, etc.). Shannon entropy $H_1 = - \sum_{i=1}^n p_i \log p_i$ characterizes a probability distribution $\{p_i\}$ on $n$ outcomes. The subscript 1 identifies Shannon entropy as the generalized R\'enyi entropy $H_\alpha$ for $\alpha = 1$~\cite{renyi}. If all $p_i$ are equal $H_1 = \log(n)$. The ``volume'' in that case is length $n$; a flat probability distribution reflects no acquired information and the effective volume is the maximum value. If only one $p_i$ is nonzero then $H_1 = \log(1) = 0$; the effective volume is minimized. Those extremes apply to all $H_\alpha$ as volume measures. The {\em difference} between two entropies, therefore the log of a volume {\em ratio},  then describes acquired information which may represent reduction of alternatives in the face of new data.  For instance, one bit of information represents the reduction of a volume of alternatives by factor 2. A single entropy by itself cannot measure information.

Spectrum data represent geometric  information in the sense that the volume defined by the data and their statistical uncertainties may be much smaller than the volume defined by an {\em a priori} hypothetical model with its uncertainties and possible systematic errors. Figure~\ref{errorrats} suggests how the data volume in the sense used above is reduced with increasing event number and particle multiplicities, thereby increasing the information relative to a fixed reference. A subset of geometric data information is {\em physically-interpretable} information that may be characterized by a model. One route to a physical model is {\em data compression} in which the original data (represented by perhaps $10^9$ particle momenta as parameters) is reduced to a few-parameter model that serves as a good representation (lossy compression) or a perfect representation (lossless compression) within statistical uncertainties.

The TCM as first reported in Ref.~\cite{ppprd} provides an example of lossy data compression. The model was empirically derived from data through a sequence of inductive steps. The main features of the 200 GeV \pp\ spectrum data were well represented but there were significant data-model deviations. The same can be said for the recent analysis of 13 TeV \pp\ data with fixed TCM in Ref.~\cite{newpptcm}. The results in Figs.~\ref{v0mfit} and \ref{spdfit} (d) show that the volume defined by the fixed-TCM hard component (represented by the dash-dotted lines) strongly overlaps but does not coincide with the data volume: the data compression is thus lossy. By allowing limited variation of hard-component parameters in Sec.~\ref{tcmfit} lossless data compression is achieved as demonstrated by Fig.~\ref{corrected} (right).

A parametrized data model may or may not be physically interpretable. An example is reported in Ref.~\cite{bayes}. An analysis of model quality (characterized by the term ``evidence'') based on Bayesian inference was applied to azimuth angular correlation data obtained from 200 GeV \auau\ collisions. The TCM applied to azimuth correlations is compared to a Fourier cosine series that has been conventionally applied to correlation data to infer elliptic flow ($v_2$) and ``higher-harmonics'' data. 
The TCM represents a jet hypothesis and is therefore predictive and falsifiable.
The Fourier series was applied with a varying number of terms and was not predictive, could accommodate any distribution on periodic variable $\phi$. 

The Bayesian analysis determined that the TCM is far superior to the Fourier series no matter what  the number of its terms or the collision centrality. The fundamental issue is the {\em predictivity} of a model. Whereas the TCM could provide a statistically acceptable data description with relatively large parameter uncertainties (strongly predictive) the Fourier series achieved an acceptable data description only with very precise parameter values representing a large amount of information derived from the data. In effect, Bayesian analysis penalizes a model to the extent that it acquires information from newly-obtained data, an elaboration of Occam's razor. The very precise Fourier coefficients are not predictive and not physically interpretable. In contrast, and as noted previously, the TCM components have been directly and quantitatively related to fundamental QCD processes that are expected in high-energy hadron-hadron collisions.

A key element of the scientific method is model falsification wherein if certain data (trusted according to community  standards) are definitively inconsistent with a model it should be discarded. In the present context one should ask: does the volume defined by a model and its parameter uncertainties significantly overlap the volume established by available trusted data. If yes the model may or may not represent the physical processes that produced the data. If  no the model is falsified.

The question of overlapping model and data volumes is quantified by Z-scores (differentially) or $\chi^2$ (integrally) which can be interpreted as measuring the ``distance'' between model and data volumes in units of data (and possibly model) uncertainties.
The fixed TCM exhibits large Z-score values for some combinations of \nch\ and event selection methods, but for others the numbers are favorable. Statistically that indicates the model volume overlaps the data volume but does not coincide with it, suggesting that the model may be adjusted to better accommodate data. Adjustment described in Sec.~\ref{tcmfit} then represents acquisition of {\em physically-relevant} information from data, one objective of data acquisition.
In contrast, the Tsallis and BW models exhibit no volume overlap with spectrum data, as demonstrated by the Z-scores in Fig.~\ref{compare} (a,b) and Fig.~\ref{bwfits} (c). The parameter variations in Figs.~\ref{qt} and \ref{dumb} nominally represent information acquisition similar to that associated with the Fourier series in Ref.~\cite{bayes}, but in contrast to the latter case the data descriptions in the former case remain unacceptable. 

\subsection{Interpreting model-parameter trends}

The TCM as derived in Ref.~\cite{ppprd} was an inductive model of spectrum data; no physical assumptions were invoked in that analysis. As such, the TCM represented a form of data compression without regard to interpretation of the algebraic components.  Subsequent to publication of Ref.~\cite{ppprd} the two components of the TCM were investigated as to physical interpretation. It was concluded that the TCM soft component results from dissociation of projectile protons after inelastic collision, and the hard component represents jet fragments from large-angle scattering of low-$x$ gluons. The latter is confirmed by quantitative prediction of spectrum hard components based on measured jet energy spectra and measured fragmentation-function ensembles~\cite{eeprd,fragevo}. The same quadratic relation between data soft and hard components is associated with measured jet energy spectra~\cite{jetspec2}. The two components of the TCM thus correspond with QCD processes that are reasonably expected in high-energy nuclear collisions. It is of interest to examine the degree of uncertainty associated with models such as Tsallis and blast-wave that, for \pp\ collisions at least, represent collision scenarios seemingly quite unlikely.

Reference~\cite{cleymans} reports the following conclusions: ``The collective radial flow velocity, $\langle \beta_t \rangle$ is almost independent of the collision energy and multiplicity classes. The kinetic freeze-out temperature, $T_{kin}$ however, shows a clear dependence on the multiplicity classes.'' It is notable that reported $\langle \beta_t \rangle$ remains constant at 0.65 down to the lowest V0M \nch\ class $n = 10$, whereas jet production is negligible there due to selection bias [see Fig.~\ref{tcm13} (a)]. What density gradient then produces the flow?
 In contrast, Ref.~\cite{qgpreview}  (a recent review of QGP-related LHC results) observes in regard to its Fig.~3 (presenting ALICE BW fit parameters) ``For small systems pp and p-Pb, the $T_{kin}$ remains constant while $\langle \beta_T \rangle$ increases rapidly with multiplicity. At similar multiplicity values, the $\langle \beta_T \rangle$ is larger for small systems.'' That summary is consistent with results reported by ALICE in Ref.~\cite{alicepppid}, Fig.~4. Reference~\cite{cleymans} summarizes: ``Conclusively, the BGBW [BW model] explains the bulk part of the transverse momentum spectra and the description is better for the higher multiplicity classes....'' But again,  the ``bulk part'' is where the great majority of jet fragments appears as the spectrum hard component which has been quantitatively related to measured jet production~\cite{eeprd,fragevo,jetspec,jetspec2}.

In terms of establishing the uncertainties for BW model-fit results the following questions arise: Does the parameter $\langle \beta_t \rangle$ actually represent a ``radial flow velocity?'' Does the parameter $T_{kin}$ actually represent a ``kinetic freeze-out temperature.'' Such terminology is conventionally associated  with particles emerging from a flowing dense medium. Is such a medium formed in \pp\ collisions? Should the mere application of a model to spectrum data impose interpretation of collision dynamics in terms of the assumptions supporting the model?

Such questions are relevant even if a model provides a statistically acceptable data description, but in the present case both the BW and Tsallis models are definitively falsified by data as in Fig.~\ref{systematic}. Aside from that evidence there have been major disagreements between different applications of the BW model to the same spectrum data as noted above. Given such problematic results it seems reasonable to reject the {\em assumptions} forming the basis for such models, whereas the assumptions inferred from a model such as the TCM that does provide an acceptable data description may be considered a reasonable working hypothesis subject to future challenges.

\subsection{Is there collectivity in \titlepp\ collisions?}

A principal motivation for the study reported in Ref.~\cite{alicenewspec} was response to recent claims that \pp\ (and \ppb) collisions exhibiting some data features identified in \aa\ collisions with QGP formation may also achieve QGP formation. Certain data features are interpreted to provide evidence for radial flow and anisotropic flows (``collectivity'') and strangeness enhancement. While Ref.~\cite{alicenewspec} included broad comparisons between data and certain Monte Carlos based on various hypothetical mechanisms there was no apparent conclusion relating results of the study to collectivity. Observations of ``qualitative'' Monte Carlo agreements or disagreements relative to various data features did not provide a definitive answer.

The main motivation for the heavy ion programs at the RHIC and LHC was based on the assumption that a large collision space-time volume would be required to insure adequate particle rescattering leading to local thermalization, hence central \aa\ collisions. Accordingly, small collision systems should not produce a QGP and should thus serve as {\em experimental controls}. On that basis \dau\ results from the RHIC were invoked to buttress QGP interpretations for certain \auau\ data features.
However, if data features identified as signaling QGP formation in \aa\ collisions are later observed in small systems (the current situation) a crisis emerges: Either QGP is a universal manifestation in all high-energy collisions or some data features were wrongly interpreted for \aa\ collisions.

The 13 TeV \pp\ \pt\ spectrum data from Ref.~\cite{alicenewspec} could respond to those questions in at least two ways: (a) Is there evidence for radial flow in \pp\ spectrum data? (b) Is there evidence for jet modification within a dense QGP medium in \pp\ spectrum data? To answer those questions requires a reliable {\em reference model}. The Monte Carlo models invoked in Ref.~\cite{alicenewspec} are insufficient in that they are not fixed (many parameters are optimized by fits to data for each collision system), and hypothetical mechanisms are complex and ill-defined (from a user perspective). In contrast, the TCM is clearly and quantitatively related to basic QCD theory and jet measurements and describes accurately a broad array of collision systems. 

The results in Ref.~\cite{newpptcm}, especially those for V0M event selection and higher \nch, already leave no room for a radial-flow contribution (source boost distribution with resulting modification of the spectrum soft component) and therefore argue against flow-related azimuthal anisotropies as {\em modulations} of radial flow.  The results in Sec.~\ref{bwmodel} of the present study demonstrate quantitatively, via {\em standard statistical measures}, how large is the distance between the BW model and \pp\ spectrum data.

Results in Sec.~III of Ref.~\cite{newpptcm}, based on a fixed TCM that accurately describes fixed modes and general shapes of isolated hard components, already leave little room for significant jet modification in \pp\ collisions. However, more-detailed analysis in Sec.~\ref{tcmfit} of the present study, based on variation of the TCM hard component, reveals quantitatively what aspects of $\hat H_0(y_t;n_{ch})$ vary with \nch\ and event selection. One may conclude that substantial event-wise fluctuations of the data hard component occur both above and below its mode. V0M event selection does not couple to fluctuations above the mode (e.g.\ TCM parameter $q$ is effectively fixed) but does bias fluctuations below the mode (parameter $\upsilon$ varies strongly). SPD event selection is complementary. Given complementary V0M and SPD $\eta$ acceptances relative to projectile longitudinal dissociation at larger $\eta$ and jet formation near midrapidity the spectrum data, via TCM analysis, suggest that jet modification is negligible; jets are consistent with conventional QCD within data uncertainties. 

It has been argued that the ``CMS ridge'' reported for high-multiplicity 7 TeV \pp\ collisions~\cite{cms} provides significant evidence for collectivity. However, the \nch\ dependence of a nonjet quadrupole (of which the ``ridge'' is one lobe) has been accurately measured for 2D angular correlations from 200 GeV \pp\ collisions~\cite{ppquad}. Whereas jet-related correlation structure varies as number of correlated pairs $\propto \bar \rho_s^2$ the quadrupole structure varies as number of pairs $\propto \bar \rho_s^3$, accurate over one order of magnitude in $\bar \rho_s$ and therefore {\em three orders of magnitude} in quadrupole amplitude. While dijet production depends quadratically on partons (low-$x$ gluons) interacting in pairs the cubic quadrupole trend suggests a three-gluon QCD interaction. There seems to be no relation to pressure gradients in a flowing dense medium. Overall, the high-statistics \pt\ spectra from Ref.~\cite{alicenewspec} support a null result for the hypothesis of collectivity in \pp\ collisions.

\section{summary} \label{summ}

The two-component (soft+hard) model (TCM) of hadron production near midrapidity in high-energy nuclear collisions was initially derived empirically from spectrum data based on two fixed \yt-dependent model functions (\yt\ being transverse rapidity). The two components were later interpreted to represent longitudinal projectile-nucleon dissociation (soft) and large-angle scattering of low-$x$ gluons with transverse fragmentation to dijets (hard). In  that version the fixed TCM is effectively a {\em predictive} model in competition with other theoretical models. The fixed TCM was recently applied to high-statistics \pt\ spectrum data from 5 and 13 TeV \pp\ collisions sorted according to two event-selection methods (V0M and SPD). Jet-related and nonjet contributions were accurately separated, and event-selection biases were determined relative to the fixed TCM reference.

One emphasis of the present study is exploration of the consequences of allowing TCM model parameters to vary over a limited range so as to best accommodate evolution of spectrum data with changes in event selection criteria (e.g.\ selection according to V0M or SPD $\eta$ acceptance). Model quality is evaluated with Z-scores, data-model differences divided by statistical uncertainties. Z-scores provide an alternative to frequently-employed data/model {\em ratios} and are the preferred measure of model-fit quality in conventional statistical analysis. The relation between spectrum ratios and Z-scores is established, revealing that spectrum ratios strongly suppress manifestations of data-model differences at lower \pt\ while exaggerating such differences at higher \pt.

The variable TCM describes spectrum data with Z-scores $O(10)$, compared with $O(100)$ for the fixed TCM. Whereas the latter reveals the form of spectrum biases due to selection bias the former provides detailed quantitative descriptions of changes in the data hard component due to selection bias. As a  result, it can be concluded that V0M selection results in changes to the hard component only below the mode of the peaked distribution, whereas SPD selection results in strong changes in the hard component only above the mode. Those results suggest that the V0M $\eta$ acceptance coincides with the splitting cascade leading to an event-wise parton distribution function (PDF) whereas the SPD acceptance corresponds to jet formation in response to low-$x$ gluon scattering.

It is observed that all spectrum data exhibit the same systematic distortion pattern, with short-wavelength structure. A method is devised to correct the distortions. With that correction r.m.s.\ Z-scores for the TCM fall in the range 3-4. Given that the remaining data-model deviations are all short-range, with no systematic pattern, it is reasonable to conclude that the variable TCM is statistically fully consistent with the spectrum data. The resulting TCM  arguably represents all information carried by the \pt\ spectra of unidentified hadrons, and the hard component has been quantitatively related to the measured properties of eventwise-reconstructed jets.

Another emphasis of this study is comparison of the variable TCM with two other spectrum models that have been frequently applied to spectrum data from the RHIC and LHC: the Tsallis model (assuming a partially-equilibrated thermodynamic system) and the blast-wave (BW) model (assuming a radially-expanding source). The comparison is based on a separate model study applied to the same 5 and 13 TeV \pp\ spectrum data.

Although the Tsallis model is fitted to spectrum data using a conventional $\chi^2$ minimization procedure the resulting Z-scores are in the hundreds.  Some of the systematic data-model deviations may result from the chosen alternative form of the Tsallis model that includes an extra factor \pt\ compared to the standard version. Tsallis model parameter $q$ increases with charge multiplicity from a value approaching the equivalent of the TCM soft component to a value approaching the hard-component equivalent. Model parameter $T$ varies from 60 to 100 MeV compared with the TCM fixed value $T \approx 145$ MeV.

The fitted BW model similarly exhibits Z-score values in the hundreds. The form of the data-model deviations is essentially what might be expected from a model assuming a boosted particle source applied to unboosted spectra. The model falls below data at lower \pt, rises above data at an intermediate point (to minimize $\chi^2$) and then drops away rapidly at higher \pt\ because the model includes no jet component. The model parameters emerging from this particular analysis differ sharply from other studies. Transverse speed $\langle \beta_t\rangle$ remains constant near 0.65 for all \pp\ event classes whereas kinetic freezeout temperature $T_{kin}$ increases by 50\%. Other studies report strongly increasing $\langle \beta_t\rangle$ and nearly constant $T_{kin}$.

Comparing the TCM with the alternative models the variable TCM Z-scores indicate statistical compatibility with the spectrum data according to standard statistical measures. The $\chi^2$ per degree of freedom is $O(10)$ which, given the complexity of generating spectra from tracking data, appears satisfactory. Both the fixed TCM and variable TCM reveal precise differential information on the response of spectrum structure to event selection methods. The evolution of model parameters for the variable TCM are physically interpretable. In contrast both Tsallis and BW models generate Z-scores consistent with $\chi^2$ per degree of freedom exceeding 10,000. The fitted parameter variations are nonintuitive (e.g.\ large $\langle \beta_t\rangle$ for \pp\ collisions with extremely low particle densities) and in some cases conflict strongly with other applications of the same model. One can then question the physical interpretation of the parameters and the relevance of the models to \pp\ spectrum data.


\end{document}